\documentclass[12pt]{iopart}
\usepackage{graphicx}
\usepackage[normalem]{ulem}
\usepackage{longtable}
\usepackage{booktabs}
\usepackage{floatrow}

\begin{document}

\title{Spin-orbit effects in pentavalent Iridates: Models and materials}

\author{Sayantika Bhowal}
\address{School of Physical Sciences, Indian Association for the Cultivation of Science, Jadavpur, Kolkata 700 032, India}
\address{current address: Materials Theory, ETH Z\"urich, Wolfgang-Pauli-Strasse 27, Z\"urich 8093, Switzerland}
\ead{sayantika.bhowal@mat.ethz.ch }

\author{Indra Dasgupta}

\address{School of Physical Sciences, Indian Association for the Cultivation of Science, Jadavpur, Kolkata 700 032, India}
\ead{sspid@iacs.res.in }

\vspace{10pt}
\begin{indented}
\item[]
\end{indented}

\begin{abstract}
Spin-orbit effects in heavy 5$d$ transition metal oxides, in particular, iridates, have received enormous current interest due to the prediction as well as the realization of a plethora of exotic and unconventional magnetic properties. While a bulk of these works are based on tetravalent iridates ($d^5$), where the counter-intuitive insulating state of the rather extended 5$d$ orbitals are explained by invoking strong spin-orbit coupling, the recent quest in iridate research has shifted to the other valencies of Ir, of which pentavalent iridates constitute a notable representative. 
In contrast to the tetravalent iridates, spin-orbit entangled electrons in $d^4$ systems are expected to be confined to the $J = 0$ singlet state without any resultant moment or magnetic response.
However, it has been recently predicted that, magnetism in $d^4$ systems may occur via magnetic condensation
of excitations across spin-orbit-coupled states. In reality, the magnetism in Ir$^{5+}$ systems are often quite debatable both from theoretical as well as experimental point of view. Here we provide a comprehensive overview of the spin-orbit coupled $d^4$ model systems and its implications in the studied pentavalent iridates. In particular, we review here the current experimental and theoretical understanding of the double perovskite ($A_2B$YIrO$_6$, $A =$ Sr, Ba, $B =$Y, Sc, Gd), 6H-perovskite (Ba$_3M$Ir$_2$O$_9$, $M =$ Zn, Mg, Sr, Ca), post-perovskite (NaIrO$_3$), and Hexagonal (Sr$_3$MIrO$_6$) iridates, along with a number of open questions that require future  investigation.  
\end{abstract}

\section{Introduction}
 
 Spin-orbit coupling (SOC) is a relativistic effect, which may be thought of as an interaction between the intrinsic spin moment of an electron and the magnetic field, generated in the rest frame of the electron due to its orbital motion around the positively charged nucleus. This interaction results in a coupling between the spin ($\vec S$) and the orbital ($\vec L$) angular momenta of an electron, lifting the $(2S+1)(2L+1)$ -fold degeneracy of an atomic energy level, and hence the name ``Spin-orbit coupling". Mathematically, the coupling emerges naturally from the Dirac Hamiltonian and may be written as ${\cal H}_{\rm SOC} = \frac{\hbar^2}{2m_e^2c^2} \frac{1}{r} \frac{dV}{dr} (\vec L \cdot \vec S) = \lambda \vec L \cdot  \vec{S}$, where the potential  $V(r) = -\frac{Ze^2}{4\pi\varepsilon_0r}$ due to the nuclear charge $Ze$ determines the coupling constant $\lambda$.

 The strong dependence of $\lambda$ on the atomic number $Z$ \cite{note,LL,Shanavas2014}, renders the inevitable spin-orbit effects in the heavy elements of the periodic table, leading to a wide variety of fascinating  phenomena, which has continued to be the center of attention in the condensed matter research over a long period of time.  This includes, for example, various topological effects \cite{Kane,Fu,Hasan,Qi,Ando}, that stem from the non-trivial geometry of the electronic Bloch bands, providing not only the platform to realize the well known particles in the standard model of relativistic high-energy physics 
  in the low energy excitations of the condensed matter systems, but also leading to the emergence of the field of {\it quantum computation}. In addition, there are Rashba and Dresselhaus spin-orbit interactions \cite{Rashba,BychkovRashba,Manchon2015} due to a gradient of electrostatic potential in non-centrosymmetric systems, which constitute the heart of the present day hot topic of {\it spin-orbitronics} research \cite{Culcer,Andrzej,Hikino,Tkachov,Greening,Zarezad,Alidoust2020,Alidoust2015,Alidoust2021}.
  The present review, however, focuses on a different aspect of the SOC effects, namely, the interplay between SOC and Coulomb interaction in a crystalline solid, which often leads to unconventional phases of matter including the novel spin-orbit Mott insulating state.

 While the importance of SOC in correlated systems had already come into notice \cite{TaSe2,TaS2,Liu2008}, the breakthrough came with the pioneering work of Kim {\it et. al}~\cite{Kim} in the year 2008, where the authors have shown that the SOC is absolutely essential to describe the insulating state of Sr$_2$IrO$_4$, as the rather moderate Coulomb interaction alone can not explain the opening of an energy gap in the extended Ir-5$d$ states. 
 As predicted by Jackeli and Khaliullin \cite{Jackeli}, this unique combination of SOC and Coulomb interaction can offer the long-awaited 
realization of the Kitaev model, that hosts a number of fascinating properties including anyonic excitations, topological degeneracy, etc., in real condensed matter systems, replacing the usual Heisenberg spin model. The idea of spin-orbit assisted Mott insulating state led to extensive studies over the past decade, in the search for new spin-orbit coupled correlated materials, their characterizations, as well as reinvestigation of previously synthesized materials in the light of the new insights on SOC.

Among these various spin-orbit coupled  systems, the family of iridates have received considerable attention, primarily because of the availability of various crystal geometries as well as different Ir valencies within the same family. 
A large section of the studied iridates share the same Ir$^{4+}:d^5$ electronic configuration as the prototypical spin-orbit Mott insulator Sr$_2$IrO$_4$, where the different connectivities of the IrO$_6$ structural unit resulting from diverse crystal geometries,  
 give rise to a plethora of exotic electronic and magnetic properties of matter such as novel spin-orbit entangled metallic state, \cite{Hirata,Panda,Sun2017,Nelson2019,Sen}, magnetic Kitaev interactions \cite{Hill,Ye,Singh2,Sizyuk,Rau,Biffin,Biffin2,Kimchi,Becker,Chun,Takayama,Catuneanu,Bhowal-BTIO2019,Aczel}, various topological states  \cite{Pesin,Wan,Carter,Ueda2012,Zhang,Sun,Ueda,Sushkov,Xu,Nelson,Bhowal2019}, e.g.,  Dirac and Weyl semimetal, nodal line semi-metals, Axion insulators, quantum spin liquid states \cite{Okamoto,Dey2012,Lee}, hidden order multipolar state \cite{Lovesey,Ganguly,Zhao,Bhowal-BTIO2019,Ganguly2020}, superconductivity \cite{Kim2016,Petrzhik}, etc.   A number of excellent reviews \cite{Krempa,Bertinshaw,Schaffer,Rau2016,Martins,Cao2018,Winter}  already exist in the literature, addressing many of these fascinating properties.

  The present review, however, focusses on another charge state of iridium, the 5+ charge state, Ir$^{5+}$ with  $d^4$ electronic configuration, known as pentavalent iridates. Similar to the tetravalent iridates, described above, pentavalent iridates also offer a wide range of structural variety. The magnetic properties of these iridates are, however, strikingly different owing to the dramatic effect of spin-orbit interaction on a local $t_{2g}^4$ electronic configuration. While in absence of SOC, degenerate $t_{2g}^4$ state results in a magnetic $S=1$ state, in presence of strong SOC orbital and spin angular momenta are coupled to form a non-magnetic $J=0$ state within an atomic limit. Locally this singlet $J=0$ ground state is separated from the higher lying triplet $J=1$ state by an energy scale that is determined by the strength of the SOC, $\lambda$. However, similar to the quantum dimer model \cite{SachdevKeimer}, the superexchange interaction between the neighboring multiplets may cause a dispersion of the triplet state, facilitating the possibility of a magnetic excitation from $J=0$ to $J=1$ state, depending on the relative values of the energy scales involved. Such a magnetic excitation maybe viewed as annihilation of singlet bosons and the simultaneous creation of triplet bosons, known as {\it triplon} excitons. The prediction \cite{Khaliullin} of such excitonic magnetism in the nonmagnetic $J=0$ ground state of $d^4$ spin-orbit coupled systems motivated tremendous interest in the pentavalent iridates.

While the first $d^4$ Ir-based material was synthesized about sixty years ago, around 1960's \cite{Figgis,Earnshaw}, the interest in the $d^4$ spin-orbit coupled materials is renewed with 
the recent advancement in the understanding of the SOC effects.
In real solids, the situation is even more complicated in the presence of additional interactions such as non-cubic crystal field ($\Delta_{\rm NCF}$), bandwidth (W), Hund's Coupling ($J_H$), Coulomb interaction ($U$), etc., which often compete among each other. Indeed, the ground state of $d^4$ iridates is determined by the delicate balance between these various competing interactions, leading to the emergence of a number of unconventional and exotic phases \cite{CaoSYIO,Bhowal2015,Nag2016,Nag2018,Nag2019}. The magnetic properties of such subtle ground state in real materials are, however, often quite debatable both from theoretical as well as experimental point of view.

The present review aims at providing an overview of the spin-orbit physics in $d^4$ iridates, taking explicitly a few notable oxide members of this family into account. The idea of this review is to focus on some of the key issues in this field as well as put forward some open questions, inspiring future work along this direction. 
We first discuss the theoretical models, developed for $d^4$ spin-orbit coupled systems, and, then, explicitly examine their implications by reviewing some of the reported $d^4$ iridates, synthesized in the form of double perovskite, post perovskite, 6H perovskite, and hexagonal structures.  Interestingly, the various crystal geometries also facilitate different connectivities of the IrO$_6$ octahedra, e.g., double pervoskite and hexagonal iridates contain isolated IrO$_6$ octahedron, while each IrO$_6$ unit in post pervoskite structure shares its edge as well corner with the neighboring octahedron. On the other hand, IrO$_6$ octahedra in 6H iridates are face-shared.  
Such diverse connectivities, in turn, provide the opportunity to investigate and compare the interplay between the different competing energy scales, governed by the specific structural geometries.

\section{Background of the $J=0$ state and Iridates} \label{background}

In this section, we provide the background of the $J=0$ state. In particular, we discuss the necessary criteria for the formation of the $J=0$ state, explain how these requirements, often, can be met in iridates, thereby, making them good candidate materials to study, and briefly mention the possibility of excitonic magnetism, which we, further, discuss in detail in section \ref{model}, which has propelled enormous interest in $d^4$ iridate research. 

\subsection{Formation of $J = 0$ state} 

\begin{figure}[t]
\centering
 \includegraphics[width=\linewidth]{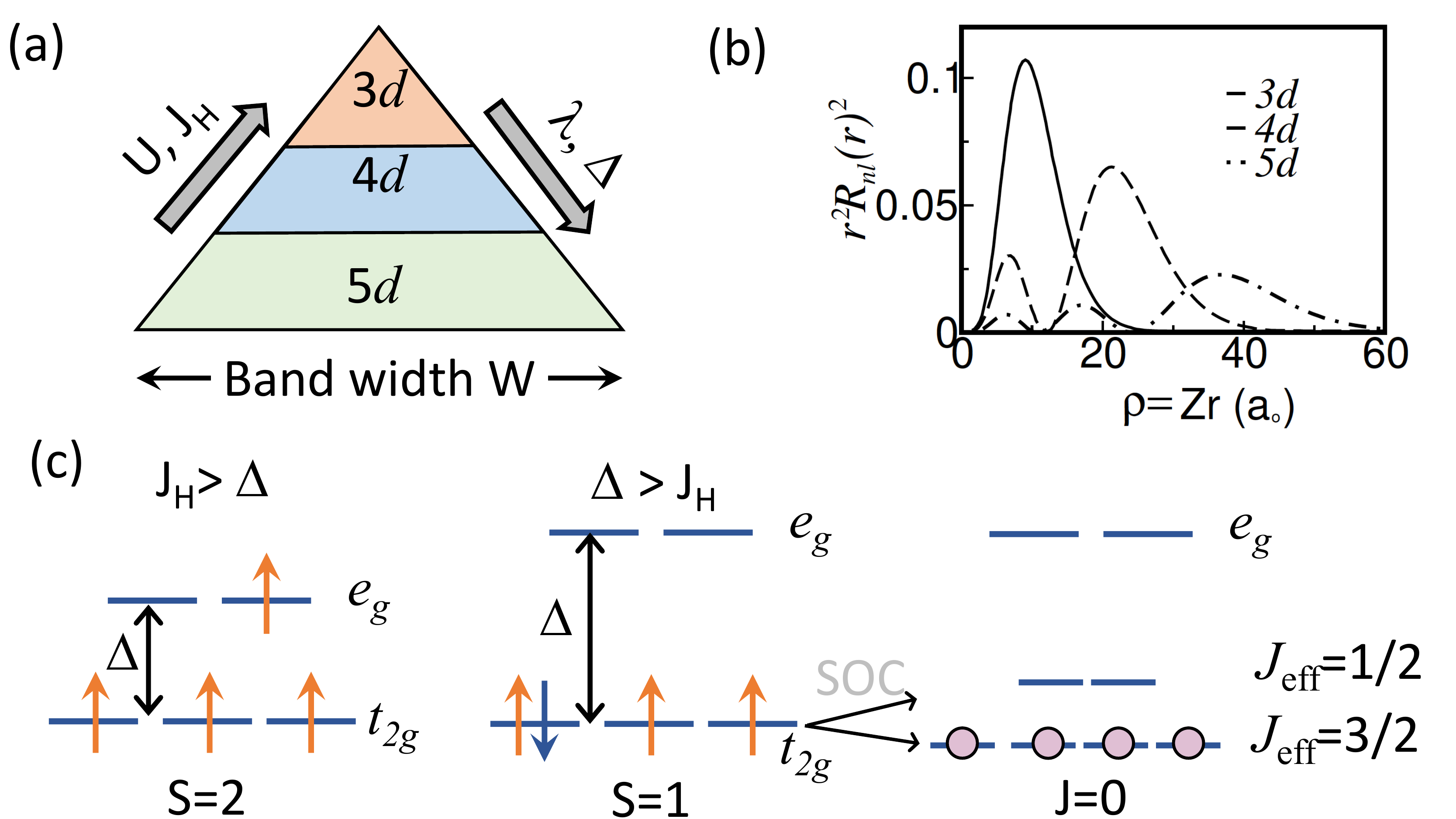}
 \caption{ The formation of $J=0$ state. (a) Schematic illustration of the different energy scales of 3$d$, 4$d$ and 5$d$ orbitals, where the base of the triangle denotes the band-width $W$ of the respective orbitals situated at an increasing height from the base. The increasing and decreasing values of the different parameters: Coulomb correlation $U$, Hund's coupling $J_H$, SOC $\lambda$, and crystal field splitting $\Delta$ along the height of the triangle are indicated by big arrows. (b) Radial distribution $r^2R_{nl}^2(r)$ for 3$d$, 4$d$ and 5$d$ orbitals  as a function of distance from the nucleus $r$, indicating that the $5d$ orbitals are more spatially extended than the rest, leading to larger $W$ as shown in (a). (c) The evolution of the ground state of a $d^4$ atomic system in an ideal octahedral environment, depending on the relative energy scales of the parameters. {\it Left:} The magnetic high spin $S=2$ state in the  $J_H > \Delta $  regime, relevant for $3d$ orbitals. {\it Middle:} The magnetic low spin $S=1$ state in the  $\Delta > J_H $ regime, typical for $4d$ orbitals in presence of weak SOC.  {\it Right:} Spin-orbit entangled non-magnetic $J=0$ state in presence of large SOC, expected for heavy $5d$ orbitals. See text for details. 
 }
\label{fig1}
\end{figure}

The primary ingredient for the formation of $J=0$ state is the strong SOC, typical for elements with large atomic number $Z$, where the electrons experience a strong nuclear potential, leading to unavoidable relativistic effect. This results in stronger SOC effects towards the end members of the periodic table. For example, in the transition metal (TM) series,  the strength of the SOC gradually increases with an added $n$ quantum number as we go down in the periodic table, viz., $\lambda_{5d} > \lambda_{4d} > \lambda_{3d}$, as also schematically illustrated  in Fig. \ref{fig1} (a).

 Interestingly, going from 3$d$ to 5$d$ TM elements, the $d$ orbitals get spatially more extended [see Fig. \ref{fig1} (b)]. As a result, the wide 5$d$ orbitals hybridize strongly with the neighboring ligand atoms, leading to a larger crystal field splitting in 5$d$ TM compounds compared to the systems with narrow 3$d$ bands. This has some interesting consequences, which play crucial role in the formation of $J=0$ state in a $d^4$ electronic configuration.  For a simple illustration, we consider TM-O$_6$ octahedral unit. Indeed such units are quite common in various transition metal oxides, particularly in iridates. We consider an ideal octahedral symmetry, which, then, splits the five-fold $d$ states into a triplet $t_{2g}$ and a higher lying doublet $e_g$ orbitals. First, let us consider the case when the TM has 3$d$ orbitals in the valence state. Due to weak crystal field splitting ($\Delta$) in these systems, the Hund's coupling $J_{\rm H}$ that favors parallel spin alignment of the electrons in different $d$ orbitals win, and $3d^4$ electronic configuration leads to a high spin $S=2$ state ($t^3_{2g}e^1_g$), as seen for example in LaMnO$_3$ \cite{Nanda}. The SOC effects being rather weak, the crystal field split states are only very weakly perturbed by SOC. In case of 4$d$ TM elements, however, the 4$d$ orbitals being wider than the 3$d$ orbitals, $\Delta$ is relatively larger in magnitude. In this case, $\Delta$ wins over the Hund's coupling $J_{\rm H}$, resulting in a low spin state $S=1$, with four electrons in the $t_{2g}$ orbitals ($t^4_{2g}e^0_g$ ) \cite{Kumar,ChenRu}. The formation of the low spin state is the first step towards the $J=0$ state. However, due to its moderate value in 4$d$ TMs, often SOC acts  only as a perturbation to the $t_{2g}$ states. It is interesting to point out here that the recent report of  a  $J=0$ ground state in 4$d$ based Ru compound K$_2$RuCl$_6$  within $LS$ coupling scheme points to the importance of SOC (beyond perturbation) also in 4$d$ based systems \cite{Takahashi}. Nevertheless, the remaining criterion of strong SOC is satisfied in $5d$ TMs, where strong SOC forms spin-orbital entangled states, destroying 
the individual orbital and spin identities of the constituents. 
 
 We pause here and discuss the formation of these new spin-orbit entangled states within a single particle formalism. 
Assuming the $t_{2g}$ states are completely decoupled from the higher lying $e_g$ orbitals, the spin-orbit entangled states can be immediately visualized by diagonalizing the SOC Hamiltonian in the $t_{2g}$ basis, viz., 
\begin{eqnarray}
{\cal H}_{\rm SOC} &= \left( \begin{array}{cc} 
   {\cal H}_+ & 0 \\
    0 & {\cal H}_-
    \end{array} \right),
\end{eqnarray}\label{soc_Ham}
where ${\cal H}_{+}$ (${\cal H}_{-}$) in the basis  $\{ |yz \downarrow\rangle$, $|zx \downarrow\rangle$, $|xy \uparrow\rangle\}$ 
($\{|yz \uparrow \rangle$, $|zx \uparrow\rangle$, $|xy \downarrow \rangle\}$) has the following form
\begin{eqnarray}
{\cal H}_{\pm} &= \left( \begin{array}{ccc} 
 0 & \mp i\frac{\lambda}{2} & \pm\frac{\lambda}{2}  \\
    \pm i\frac{\lambda}{2} & 0 & +i\frac{\lambda}{2} \\
     \pm \frac{\lambda}{2} & -i\frac{\lambda}{2} &  0
      \end{array} \right).
\end{eqnarray}\label{soc_Ham} 
We have assumed, here, degenerate $t_{2g}$ states in an ideal octahedral environment. 
The diagonalization of ${\cal H}_{\rm SOC}$ yields four-fold degenerate $\Gamma_8$ (-$\frac{\lambda}{2}$) and two-fold degenerate $\Gamma_7$ ($\lambda$) states,  where each of the eigenstates of ${\cal H}_{+}$ has its time-reversal counterpart in the corresponding eigenstate of ${\cal H}_{-}$, forming the Kramer's doublet.  The explicit form of these eigenstates are given by:
\begin{eqnarray*}
   |1\alpha\rangle &=& -\frac{1}{\sqrt{2}}(|d_{yz\uparrow}\rangle+i|d_{zx\uparrow}\rangle),\\
   |1\beta\rangle &=& \frac{1}{\sqrt{2}}(|d_{yz\downarrow\rangle}-i|d_{zx\downarrow}\rangle),\\
   |2\alpha\rangle &=& \frac{1}{\sqrt{6}}(2|d_{xy\uparrow}\rangle-|d_{yz\downarrow\rangle}-i|d_{zx\downarrow}\rangle),\\
   |2\beta\rangle &=& \frac{1}{\sqrt{6}}(2|d_{xy\downarrow}\rangle+|d_{yz\uparrow}\rangle-i|d_{zx\uparrow}\rangle),
 \end{eqnarray*}
and,
 \begin{eqnarray*}
  |3\alpha\rangle &=& \frac{1}{\sqrt{3}}(|d_{xy\uparrow}\rangle+|d_{yz\downarrow}\rangle+i|d_{zx\downarrow}\rangle),\\
  |3\beta\rangle &=& \frac{1}{\sqrt{3}}(|d_{xy\downarrow}\rangle-|d_{yz\uparrow}\rangle+i|d_{zx\uparrow}\rangle).
 \end{eqnarray*} 
 Note that the eigenstates above do not have pure spin or orbital characters any more, as they couple to form entangled states.  In the literature, these quartet and doublet  are commonly referred to as $j_{\rm eff}$~=~3/2 and $j_{\rm eff}$~=~1/2 states respectively. It is important to point out that these spin-orbit entangled effective states are obtained within a single particle formalism and they are different from the many-body states $J =1/2$ or $3/2$. To make this distinction clear, now on we represent the single particle states by $j$, while the many-body states are indicated as $J$. 
    
If we now go ahead and fill up these spin-orbit coupled states with four electrons, present in the 5$d$ shell, we can see that the lower lying $j_{\rm eff}$~=~3/2 state gets completely occupied while the  $j_{\rm eff}$~=~1/2 state remains empty. As there is no unpaired electron left in the system, the system becomes non-magnetic. The effective total angular momentum, $j_{\rm eff}$ for this state may be determined from the effective orbital angular momentum $l_{\rm eff} = -1$ (see Appendix A for the illustration of this point) and the spin angular momentum $s=1$, which again give a non-magnetic $j_{\rm eff}=l_{\rm eff} + s = 0$ state. 
The evolution of the ground state of a $d^4$ system in the different regimes of the competing energy scales is schematically shown in Fig. \ref{fig1} (c). 
 Note that the formation of $J=0$ state within the many-body framework can similarly be described assuming $t_{2g}$ only model, which we discuss in detail later in the next section.

\begin{figure}[t]
\centering
\includegraphics[scale=.2]{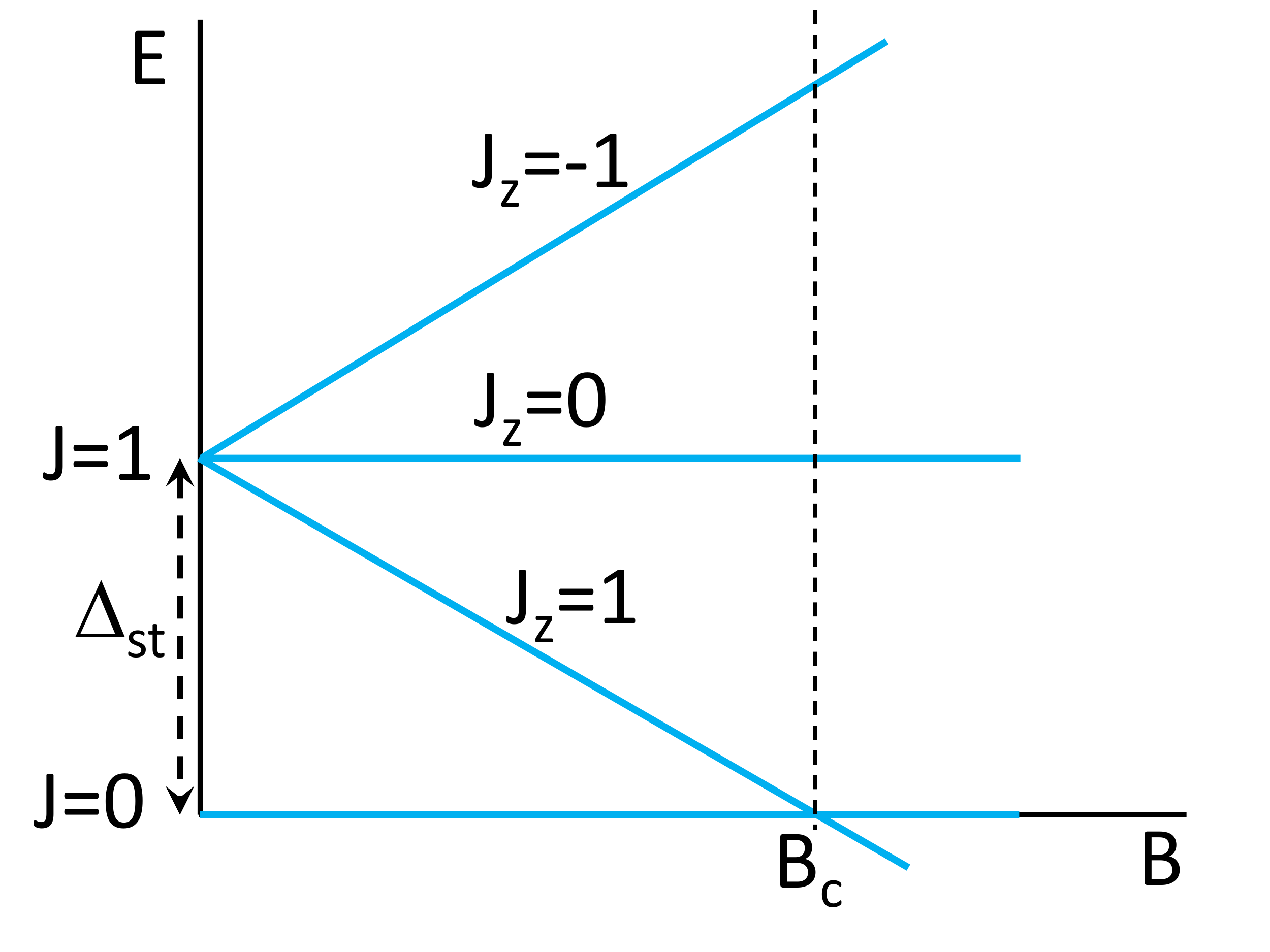}
 \caption{ The simplified picture of the magnetic-response of a singlet-triplet system in presence of a magnetic field $\vec B$. In absence of a magnetic field, $\vec  B=0$, the singlet $|J=0 \rangle$ and the triplet $|J=1 \rangle$ states are separated by an energy gap of $\Delta_{st}$, while in presence of a magnetic field the triplet state $|J=1 \rangle$ undergoes a Zeeman splitting, leading to splitting of the $|J_z =\pm 1\rangle$ states, with energies $E_{\mp}=\Delta \mp g_J \mu_B B$. When the strength of the applied magnetic field $B$ exceeds a critical value $B_c=\Delta/g_J \mu_B$, the $|J_z = +1\rangle$ state lies below the $J=0$ state, leading to a sharp magnetic transition. In reality, the behavior of the singlet state depends on the matrix elements of the magnetic dipoles between the non-magnetic ground state and the magnetic excited state. 
 }
\label{fig0}
\end{figure}

As an aside, it is interesting to point out that for five electrons in the 5$d$ shell, the  $j_{\rm eff}$~=~1/2 state gets partially occupied in addition to the completely filled $j_{\rm eff}$~=~3/2 state. This half-filled $j_{\rm eff}$~=~1/2 state is further split into empty upper and filled lower Hubbard bands in presence of Coulomb interaction, leading to the well known  $j_{\rm eff}$~=~1/2 Mott insulating state \cite{Kim}. In contrast to the case of four electrons, discussed above, here the system remains magnetically active. The role of SOC, here, is to just convert the exchange interactions between spin and orbital angular momentum into an effective total angular momentum. These interactions between the total angular momenta, however, may have  interesting consequences, for example they could be bond dependent, inherited from the corresponding interactions between orbital angular momentum for the $t_{2g}$ orbitals \cite{Khaliullin2002}. 

\subsection{Origin of excitonic magnetism in spin orbit coupled $d^4$ systems}

 The non-magnetic $J=0$ ground state, imposed by SOC, is very typical for Van-Vleck type paramagentism, where the admixture of the ground state  $|\psi_0 \rangle$ with the magnetic excited states $|\psi_n \rangle$ gives rise to a magnetic response in presence of a magnetic field ($\vec B$). In the language of perturbation theory, this can be expressed as 
$|\psi \rangle = |\psi_0 \rangle + \sum_n \frac{\langle \psi_n| (\vec B \cdot \vec J)|\psi_0 \rangle}{E_n-E_0}$ \cite{KhomskiiTMO}.  The essence of the behavior of a singlet-triplet system in presence of a magnetic field is schematically illustrated in Fig. \ref{fig0}.

 \subsubsection{Why Iridates?} 
 
 The 5$d$ TM oxides, of which iridates are notable members, facilitate the basic requirements for  the formation of a non-magnetic $J= 0$ state. In such a non-magnetic state, however, magnetism may appear by a mechanism very similar to Van-Vleck paramagentism via excitation to the higher lying triplet $J=1$ state, as suggested by Khaliullin \cite{Khaliullin}. The possibility of such excitation is governed by the relative energy scales of the singlet-triplet splitting ($\Delta_{st}$), determined by the SOC constant $\lambda$, and the superexchange energy, which is $\sim 4t^2/U$. When these two energy scales are comparable, singlet-triplet Van Vleck transitions may occur, resulting in a magnetism in the apparently non-magnetic $J=0$ state.

 \subsubsection{Singlet-triplet transition: Excitonic magnetism:}
 
\begin{figure}[t]
\centering
\includegraphics[scale=.25]{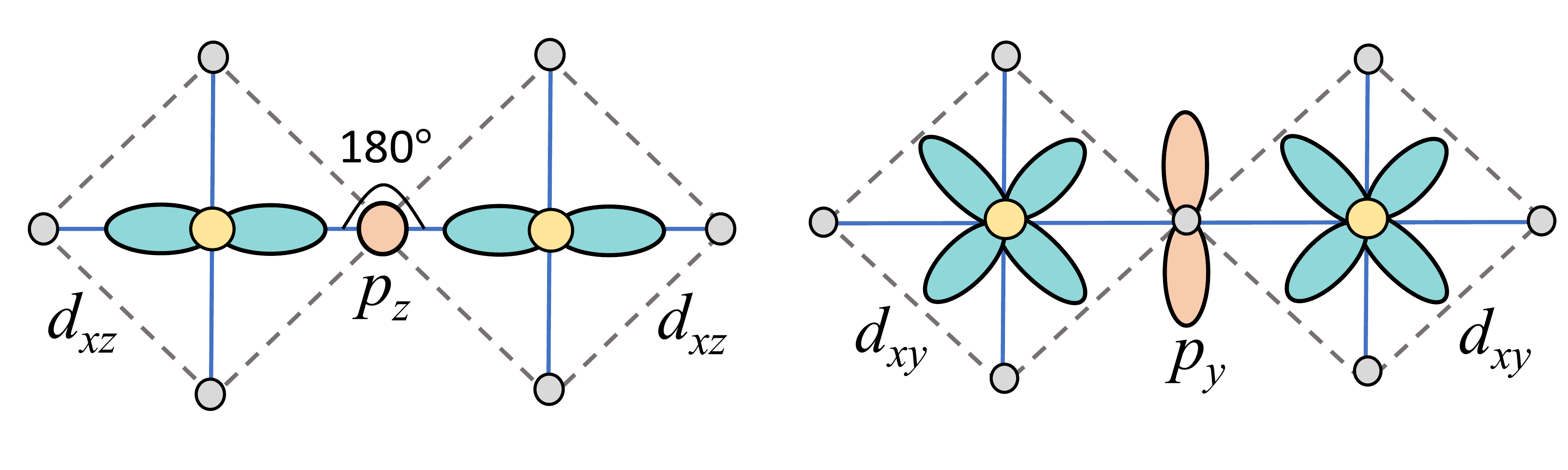}
 \caption{ The schematic illustration of the 180$^\circ$ $d-p-d$ bond.
 The figure shows the two active orbitals, $d_{xz}$ and $d_{xy}$, along a bond in the $x$-direction for corner shared octahedral geometry. 
 }
\label{fig3}
\end{figure}

Following Khaliullin's idea \cite{Khaliullin} of singlet-triplet transition, here, we briefly illustrate this process considering 180$^\circ$ $d-p-d$ bond geometry, which is relevant for the corner-shared octahedra. 
In the 180$^\circ$ $d-p-d$ bond geometry, as shown in Fig. \ref{fig3}, only two of the three $t_{2g}$ orbitals are active along a certain bond, while the third orbital, say $c$, remains inactive. For simplicity, we can call this bond $c$. Therefore, the nearest neighbor hopping matrix along this $c$ bond is given by $t(a_i^\dagger a_j+b_i^\dagger b_j + H.c.)$, indicating that the hopping is diagonal in the active orbital space. Here $a_i^\dagger (a_i), b_i^\dagger (b_i)$ are the creation (annihilation) operators that create (annihilate) an electron in the respective active $t_{2g}$ orbitals $a$ and $b$ at site $i$. In the illustration of Fig. 3, these active orbitals are $d_{xz}$ and $d_{xy}$ orbitals for the bond along $x$. We now introduce three different operators $A, B,$ and $C$, which represent the three possible orbital configurations for the $t_{2g}^4$ shell, viz., $A=\{ a^2bc\}, B=\{ ab^2c\},$ and $C=\{ abc^2\}$. The one-to-one correspondence between the operators and the orbitals are apparent from this definition. With the help of these operators, we can also rewrite the components of the orbital angular momentum operator, e.g., $L_x =-i(B^\dagger C-C^\dagger B)$, etc. 

In the strong coupling limit ($U >> t, \lambda, J_H$), this results in a spin-orbital Hamiltonian, having the form $H=(t^2/U) \sum_{\langle ij \rangle} [ (\vec S_i \cdot \vec S_j + 1) O^{(\gamma)}_{ij} +(L_i^\gamma)^2 + (L_j^\gamma)^2 ]$, where $O^{(\gamma)}$ is the bond dependent orbital operator which can be written in terms of the $A, B, C$ operators or directly expressed in terms of the orbital angular momentum operators $L_x, L_y$, and $L_z$. For example, the operator for the bond $c$ can be written as
$O^{(c)} = (L^x_i L^x_j)^2 + (L^y_i L^y_j)^2+L_i^xL_i^yL_j^yL_j^x+L_i^yL_i^xL_j^xL_j^y$. Similarly, the operators $O^{(a)}, O^{(b)}$ for other bonds $a$, and $b$ can be obtained from cyclic permutations of $L_x, L_y,$ and $L_z$.
Clearly, the Hamiltonian $H$ operates in the $(M_L, M_S)$ basis. We now want to project it onto the low-energy subspace of the singlet $J=0$ ground state, $|0\rangle = \frac{1}{\sqrt{3}} (|1,-1\rangle -|0,0 \rangle + |-1,1\rangle)$ and the first excited state $J=1$ triplet at an energy $\Delta_{st}$ determined by $\lambda$, $|T_0\rangle = \frac{1}{\sqrt{2}} (|1,-1\rangle -|-1,1 \rangle), |T_{\pm1}\rangle = \pm \frac{1}{\sqrt{2}} (|\pm1, 0\rangle -|0,\pm1 \rangle)$. Calculating the matrix elements of the orbital and spin angular momentum operators in this Hilbert space, we can write them in terms of the  ``triplon" $\vec T$ with the Cartesian components $T_x= \frac{1}{i\sqrt{2}} (T_1-T_{-1}), T_y= \frac{1}{\sqrt{2}} (T_1+T_{-1}),$ and $T_z=iT_0$, and the ``spin" $\vec J =-i (\vec T^\dagger \times \vec T)$.
For example, $\vec S =-i \sqrt{\frac{2}{3}} (\vec T -\vec T^\dagger)+\frac{1}{2} \vec J$, and $\vec L =i \sqrt{\frac{2}{3}} (\vec T -\vec T^\dagger)+\frac{1}{2} \vec J$. Using these relations we can immediately write down the  magnetic moment for a $t_{2g}^4$ configuration in terms of $\vec T$ and $\vec J$, viz.,  
$\vec M = 2\vec S-\vec L= -i\sqrt{6} (\vec T -\vec T^\dagger) +g_J \vec J $, where $g_J=\frac{1}{2}$. Clearly, the magnetism in a $J=0$ state is driven solely by the $\vec T$ exciton as the second term in $\vec M$ does not contribute for a singlet state, $J=0$, and hence, the magnetism is referred to as ``excitonic magnetism".  

The transformation $H(L,S) \rightarrow H(\vec T, \vec J)$, as described above, gives us the desired effective singlet-triplet model \cite{Khaliullin},
\begin{equation}
 H_{\rm eff} = \Delta_{st} \sum_i T_i^\dagger T_i + \kappa \frac{2}{9} \sum_{ij} [\vec T_i^\dagger \cdot \vec T_j -\frac{7}{16} (\vec T_i \cdot \vec T_j + H.c.)].
\end{equation}
Here 
$\kappa$ denotes the exchange interaction $4t^2/U$ and terms upto quadratic in $\vec T$ bosons are considered. The three and four-boson interaction terms, that include the quadrupole operators and bi-quadratic Heisenberg exchange couplings respectively, have only small values near the phase transition for 180$^\circ$ bond geometry and are, therefore, not important for the qualitative understanding of magnetic phase transition driven by the condensation of $\vec T$ bosons. The diagonalization of this effective Hamiltonian gives the energy dispersion for the triplet excitation, which is useful to understand the condensation of the triplet state. For example, when $\kappa < \kappa_c$, where $\kappa_c$ is some critical value of $\kappa$, the magnetic excitation has a finite gap $\lambda \sqrt{1-(\kappa/\kappa_c)}$, indicating a paramagnetic phase. 
At $\kappa = \kappa_c$ this gap closes and a magnetic phase transition takes place due to condensation of the $\vec T$ bosons.

 Interestingly, soon after this proposal of excitonic magnetism, Cao {\it et. al} \cite{CaoSYIO} experimentally showed the existence of magnetism  in the $d^4$ double perovskite iridate Sr$_2$YIrO$_6$. Indeed, in most of the pentavalent iridates, the Ir atom has a non-zero  effective magnetic moment (see, for example, the list of pentavalent iridates with their magnetic and trasport properties in Table \ref{tab1}), where the moment values are often attributed to the non-cubic crystal field, bandwidth effect, etc. in the solid causing the deviation from the atomic like picture, described above for the formation of $J =0$ state. Furthermore, since the strength of the SOC is renormalized in a solid as $\lambda/2S$ \cite{KhomskiiTMO,Whangbo}, where $\lambda$ is the atomic SOC, and $S$ is the total spin, one might anticipate the SOC strength would be weaker in $d^4$ systems with $S=1$ as compared to tetra-valent iridates with $S=1/2$. Such weak SOC tends to favor excitonic magnetism in $d^4$ iridates as the relative strength of the SOC constant and the superexchange energy scale determines the occurrence of excitonic magnetism. Interestingly, a recent theoretical study \cite{Kaushal} based on density matrix renormalization also suggests spin-orbit induced transition to an excitonic anti-ferromagnetic condensate even at an intermediate Coulomb interaction regime.

 The magnetism in pentavalent iridates are, however, not beyond doubt. The situation is, in particular, controversial and flooded with conflicting results for the double peroveskite iridate  both in theory and experiments \cite{Bhowal2015,Nag2018,TDey, Kunes,Hammerath,Chen,Fuchs,Gong,Terzic}. Nevertheless, the possibility of the breakdown  of the anticipated $J =0$ state, leading to the unconventional magnetism, have generated enough interest in pentavalent iridates, which are further accelerated by the observation of novel ``spin-orbital liquid" state in the 6H perovskite iridate \cite{Nag2016}. The easy availability of the $d^4$ electronic configuration in iridates, compared to other 5$d$ TM oxides, have further boosted the search for the realization of $J=0$ state and/or the excitonic magnetism in Iridates. 
\\


  \floatsetup[longtable]{LTcapwidth=table}
 
\begin{longtable}[ht]{ c c c c c c c} 
\caption{Structural, transport and magnetic properties of pentavalent iridates.  The notations FM, AFM and SOL refer to ferromagnetic, anti-ferromagnetic and spin-orbital liquid states
 respectively. The connectivity of IrO$_6$ unit in each structural family is given within the parenthesis. } \label{tab1} \\
\hline
 Crystal  \ &  Compound \ &  Space  \ &  Transport  \ & Magnetic  \ & $\mu_{\rm eff}/$Ir & Ref.\cr                           
 structure  \ & 					\ &  group  \ & property  \ & ordering \ & ($\mu_B$) & \cr
\hline
\endfirsthead
\multicolumn{7}{c}%
{\tablename\ \thetable\ -- \textit{Continued.}} \\
\hline
 Crystal  \ &  Compound \ &  Space  \ &  Transport  \ & Magnetic  \ & $\mu_{\rm eff}/$Ir & Ref.\cr                           
 structure  \ & 					\ &  group  \ & property  \ & ordering \ & ($\mu_B$) & \cr
\hline
\endhead
\hline \multicolumn{7}{r}{\textit{Continued on next page}} \\
\endfoot
\hline
\endlastfoot
 							& Sr$_2$YIrO$_6$ & $P2_1/n$ & Insulator & AFM & 0.91 & \cite{CaoSYIO,Bhowal2015}  \\
 						    &                              &  $Fm\bar{3}m$    &       Insulator  & None & 0.21 & \cite{Corredor} \\
                                & Ba$_2$YIrO$_6$ &    $Fm\bar{3}m$            & Insulator   & None & 0.3 & \cite{Bhowal2015,Nag2018} \\ 
                				&              				  &			   & 	Insulator & None & 0.44 & \cite{TDey}-\cite{Gong} \\
                				&                              & $P2_1/n$           & -  & - & - & \cite{Wakeshima} \\
    %
         Double           &                               &                &    Insulator & AFM & 1.44 & \cite{Terzic} \cr
perovskite 	        & Sr$_2$GdIrO$_6$ &    $Pn$-$3$ & Insulator    &  FM   & - & \cite{CaoSYIO,Bhowal2015} \\
     (Isolated)                            & Sr$_2$ScIrO$_6$  &   $P2_1/n$          &     Insulator    & None   & 0.16 & \cite{Kayser, Chakraborty2019, Bhowal2020} \\
                          & Sr$_2$LuIrO$_6$  &    $P2_1/n$         &     -   &None    &- & \cite{Wakeshima} \\   
                                &  Ba$_2$ScIrO$_6$ &     $Fm\bar{3}m$        &     Insulator    & None   & 0.39 & \cite{Kayser, Chakraborty2019} \\    
                                &  Ba$_2$LuIrO$_6$ &         $P2_1/n$    &     -   & None    & - &\cite{Wakeshima} \\ 
                                &   Ba$_2$LaIrO$_6$ &        $P2_1/n$   &    -    & None   & - & \cite{Wakeshima} \\ 
                                &   							 &        $R\bar{3}$   &    -    &-   & - & \cite{Fu2005} \\
                                &	Ba$_2$NdIrO$_6$ 	&        Cubic   &    -    &-   & - & \cite{Thumm} \\
                            &		Ba$_2$SmIrO$_6$	&      Cubic   &    -    &-   & - & \cite{Thumm}\\
                            &		Ba$_2$DyIrO$_6$	&       Cubic   &    -    &-   & - & \cite{Thumm}\\ 
                             &  Sr$_2$InIrO$_6$ &         $P2_1/n$    &     Insulator   & None    & 1.608 &\cite{Laguna-Marco}\\
               &  Sr$_2$CoIrO$_6$ &         $I2/m$    &     Insulator   & AFM    & 0.47 &\cite{Narayanan,Kolchinskaya,stefano}\\                
&  Sr$_2$FeIrO$_6$ &         $I2/m$    &     Insulator   &  AFM    & - &\cite{Laguna-Marco} \\ 
&  &         $I\bar{1}$    &     Insulator   &  AFM    & - &\cite{Kharkwal} \\
&  Ca$_2$FeIrO$_6$ &         $I\bar{1}$    &     Insulator   & AFM    & - &\cite{Kharkwal} \\
    & La$_2$LiIrO$_6$ &         $Pmm2$    &     -   & None    & 1.42 &\cite{Battle,Hayashi} \\                 
&  BaLaMgIrO$_6$ &         Pseudocubic    &     -   & None   & - &\cite{Battle} \\
& Bi$_2$NaIrO$_6$ &       $P2_1/n$       &     Insulator   & None   & 0.19 &\cite{Prasad} \\              
& SrLaMgIrO$_6$ &       $P2_1/n$       &     Insulator   & None   & 0.61 &\cite{Wolff} \\  
& SrLaZnIrO$_6$ &       $P2_1/n$       &     Insulator   & None   & 0.46 &\cite{Wolff}\\\  
& SrLaNiIrO$_6$ &       $P2_1/n$       &     Insulator   & AFM  & 0.19 &\cite{Wolff}\\ 
\mr             
Post perovskite & NaIrO$_3$  & $Cmcm$   &  Insulator &   NM  & 0 & \cite{Bhowal2015,Bremholm,Du} \cr
(Corner \& & & & & & & \\
 edge shared) & & & & & &  \\
\mr
 & Ba$_3$ZnIr$_2$O$_9$ & $P6_3/mmc$ & Insulator &  SOL & 0.2 & \cite{Nag2016} \\
6H-perovskite& Ba$_3$MgIr$_2$O$_9$ & $P6_3/mmc$ & Insulator &  None &  0.5-0.6 &\cite{NagBMIO2018,Sakamoto} \\
(Face shared)& Ba$_3$CaIr$_2$O$_9$ & $C2/c$ & Insulator &  None & - & \cite{NagBMIO2018,Sakamoto}  \\
& Ba$_3$SrIr$_2$O$_9$ & $C2/c$ & Insulator &  None & - & \cite{NagBMIO2018,Sakamoto}  \\
& Ba$_4$LiIr$_3$O$_{12}$ & $P6_3/mmc$ & Insulator &  Magnetic & - & \cite{Gore1996} \\
& Ba$_4$NaIr$_3$O$_{12}$ & $P6_3/mmc$ & Insulator &  -& - & \cite{Gore1996} \\
& Ba$_3$CdIr$_2$O$_{9}$ & $C2/c$ & Insulator &  None & 0.3 & \cite{Salman} \\
\mr
   Hexagonal           & Sr$_3$NaIrO$_6$ & $R\bar{3}c$ & Insulator   & None& -& \cite{Ming,Segal} \\
 (Isolated)                            & Sr$_3$LiIrO$_6$ & $R\bar{3}c$ & Insulator   & None& -& \cite{Ming,Segal,Davis} \\
 \mr          	 
Pyrochlore & Cd$_2$Ir$_2$O$_7$ & FCC & -& -& -& \cite{Sleight} \\
(Corner shared)				 & Ca$_2$Ir$_2$O$_7$ & FCC & -& -& -& \cite{Sleight} \\
\mr	
Others         & KIrO$_3$ & $Pn3$ & -&- & 1.04  & \cite{Hoppe} \\
          & Ba$_{0.5}$IrO$_3$ & $I23$  & Metal &-&  -& \cite{Sleight} \\
          & Sr$_{0.5}$IrO$_3$ & BCC & -& -& -& \cite{Sleight} \\	
          & Na$_3$Cd$_2$IrO$_6$ & $C2/m$  & -& -& -& \cite{Frenzen} \\	      	 
\end{longtable}

 \section{Microscopic model and breakdown of J=0 state} \label{model}
 
The various comparative energy scales, e.g.,  $\Delta$, $J_{\rm H}$ and $\lambda$ as discussed above, in 5$d$ TM oxides enrich the physics of $d^4$ iridates. The presence of magnetism in different iridates out of apparently non-magnetic individual Ir atoms piqued enormous scientific curiosity in the $d^4$ systems. While the possibility of excitonic magnetism has already been proposed, the understanding of the role of different competing parameters can provide the answer to
 the two important questions, that are of primary interest,  namely,  which parameter (or parameters) might drive the uncoventional magnetism in $d^4$ spin-orbit coupled systems at the first place. Secondly what would be the nature of this emergent magnetism. Generic theoretical models, both atomic as well as beyond atomic limit, developed in this context, are proven to be useful to gain valuable insight into the physics of the $d^4$ spin-orbit coupled systems. Here we give a brief overview of these results that essentially provide the ground to analyze the magnetic response of the real materials.

 \subsection{Atomic model and non-cubic crystal field effect}
\begin{figure}[t]
\centering
\includegraphics[scale=.25]{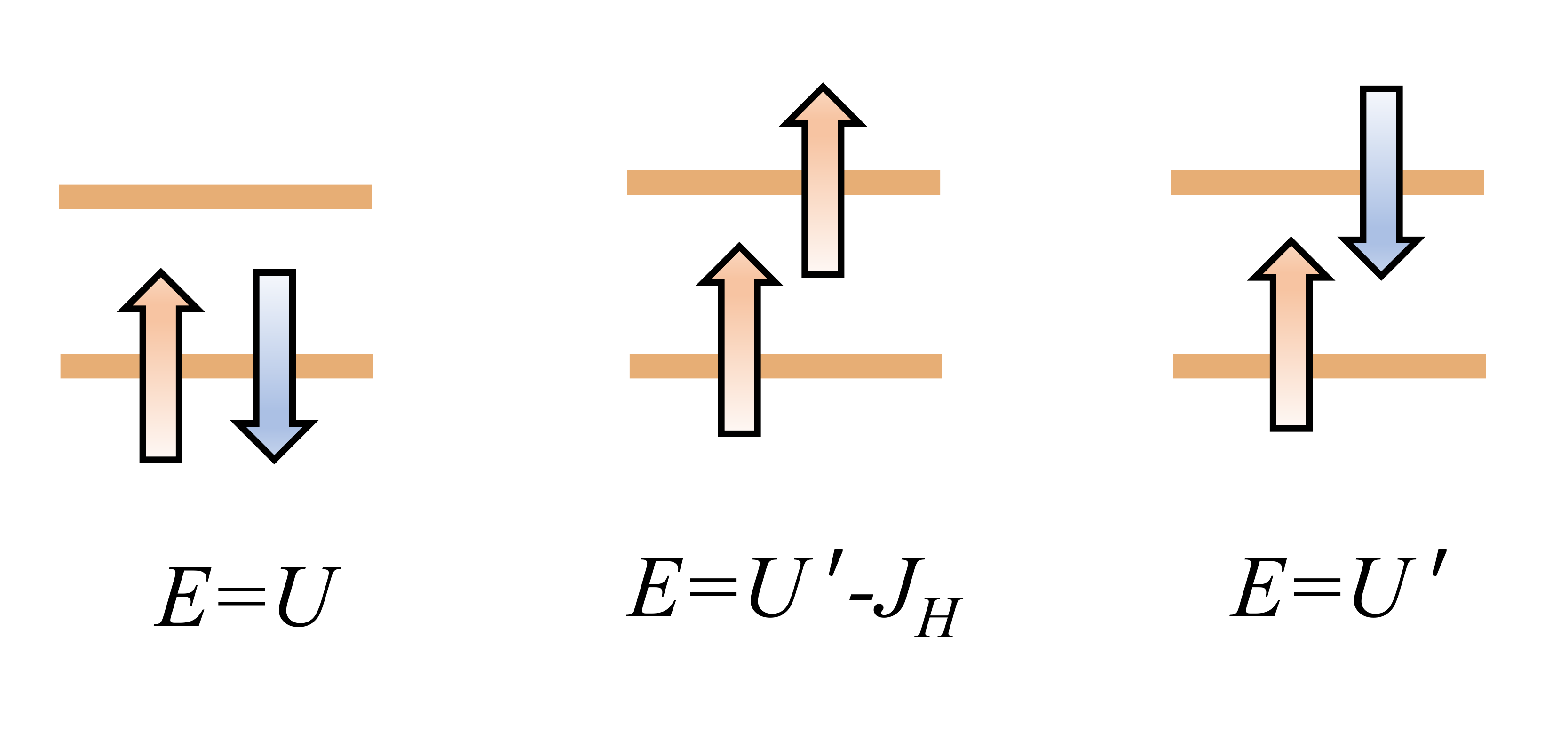}
 \caption{ Schematic illustration of the Kanamori interaction terms, intra-orbital Coulomb interaction $U$ ({\it left}), Hund's Coupling $J_H$ ({\it middle}), and inter-orbital Coulomb interaction $U'$ ({\it right}). For simplicity two orbitals and two electrons are considered, the different arrangements of which leads to different energies $E$ depending on the various interactions. 
 }
\label{fig-new}
\end{figure}

For simplicity, we first consider a many-body atomic model and, then, eventually, we shall include the band structure effects by allowing hopping between the spin-orbit coupled atoms. 
Assuming the $e_g$ orbitals are completely decoupled from the $t_{2g}$ sectors owing to the strong octahedral splitting $\Delta$, only the $t_{2g}$ orbitals are considered at each atomic site. The corresponding atomic many-body Hamiltonian, spanned in the Hilbert space of dimension $^6C_4 = 15$, includes the electron-electron interactions as well as the SOC interaction, viz.,  
\begin{equation}\label{AH}
  {\mathcal H}_{\rm atom}~=~ {\mathcal H}_{int} + {\mathcal H}_{SO}.
 \end{equation}
Here the Kanamori Hamiltonian ${\mathcal H}_{int}$ contains the intra-orbital and the inter-orbital Coulomb interactions $U$ and $U^\prime$, which are related to each other via the Hund's coupling $J_H$,viz., $U~=~U'+2J_H$. The schematic illustration of the various interaction terms are given in Fig. \ref{fig-new}. The explicit form of the Hamiltonian at an atomic site $i$ is given by \cite{Bhowal2020,Matsuura,Matsuura2014,Bhowal2018}
 \begin{eqnarray}\nonumber
   {\mathcal  H}^i_{int} &=& U \sum_{l} n_{i,l \uparrow}n_{i,l \downarrow} 
        + \frac{U^{\prime}-J_H}{2} \sum_{l \neq m,\sigma}                    n_{i,l\sigma} n_{i,m \sigma} 
        + \frac{U^{\prime}}{2} \sum_{\sigma \neq \sigma^{\prime}}\sum_{l \neq m}                                                 n_{i,l\sigma}n_{i,m\sigma^{\prime}} \\ \nonumber
         & -& \frac{J_H}{2} \sum_{l \neq m}
        (d_{i,m \uparrow}^\dagger d_{i,m \downarrow}d_{i,l \downarrow}^\dagger d_{i,l \uparrow}+         d_{i,m \uparrow}^\dagger d_{i,m \downarrow}^\dagger d_{i,l \uparrow} d_{i,l \downarrow}+
        h.c.)
\end{eqnarray}
and the spin-orbit interaction is 
 \begin{eqnarray}
{\mathcal  H}^i_{SO} &=& \frac{{\dot{\iota}} \lambda}{2} \sum_{lmn} \epsilon_{lmn} \sum_{\sigma \sigma'} \sigma_{\sigma \sigma^\prime} ^n d_{i,l\sigma}^\dagger d_{i,m\sigma'}.
\end{eqnarray}
Here, $i$-$l$-$\sigma$ denotes the site-orbital-spin index and $d_{i,l\sigma}$($d_{i,l\sigma}^\dagger$) is the annihilation (creation) operator.

\begin{figure}[t]
\centering
 \includegraphics[scale=.3]{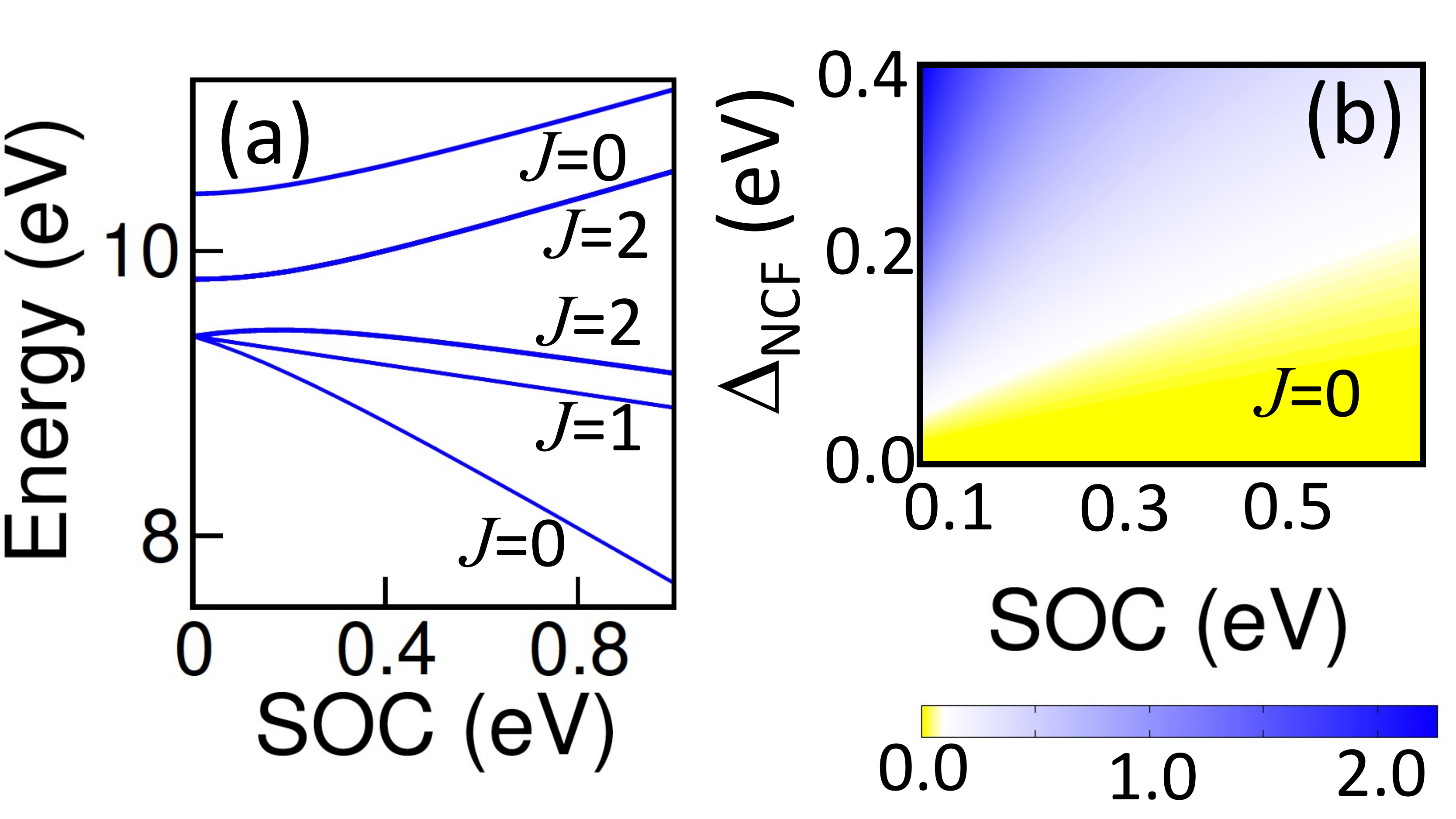}
 \caption{Emergence of magnetism in a spin-orbit coupled atomic $d^4$ model. (a) The energy evolution of the many-body states of the atomic model, Eq. \ref{AH},  as a function of SOC for the $d^4$ electronic configuration. The ground state of the atomic model is a non-magnetic $J=0$ state. (b) The emergence of magnetism in the phase diagram of an atomic model  in presence of non-cubic crystal field $\Delta_{\rm NCF}$, showing the variation of the ground state expectation value of $J^2$ operator in the parameter space of SOC $\lambda$ and $\Delta_{\rm NCF}$ for a fixed value of $J_H/U =$ 0.1. $\Delta_{\rm NCF}$ in (b) represents the energy separation between the subsequent $t_{2g}$ levels, the degeneracy of which is broken completely and are energetically equispaced. The figure is reprinted from Ref. \cite{Bhowal2020}, \copyright 2020, with permission from Elsevier.
 } 
\label{fig2}
\end{figure} 
 
The exact diagonalization of the atomic model, Eq. \ref{AH}, gives the spin-orbit coupled eigenstates that are characterized by the total angular momentum quantum number $J$: $J = 0$ (1), $J = 1$ (3), $J = 2$ (5), $J = 2$ (5), and $J = 0$ (1) with increasing order in energy [see Fig.\ref{fig2} (a)]. Here, the number within the parenthesis denotes the degeneracy of the states. 
As anticipated, $J=0$ singlet forms the ground state (similar to the single particle picture, described in the previous section ) which is separated from the excited triplet $J=1$ state.
The energies of these states are dictated by the interaction parameters such as Hund's coupling $J_H$, Coulomb interaction $U$ and SOC constant $\lambda$, while their energy difference is independent of $U$. The exact energy eigenvalues of the atomic states $J = 0$ and $J = 2$ depend on the relative strength of $J_H$ and $\lambda$ (In the limit $J_H >> \lambda$, $E_{J=0} = E_0 - 4U + 7J_H - \lambda$ and $E_{J=2} = E_0 - 4U + 7J_H + \lambda/2 $; for $J_H << \lambda$, $E_{J=0} = E_0 - 4U + 7J_H - 2\lambda$ and $E_{J=2} = E_0 - 4U + 7J_H - \lambda/2$) but their energy difference is always  $3\lambda/2$. The energy eigen value of the $J=1$ state, on the other hand, always have the same analytical form $E_{J=1} = E_0 - 4U + 7J_H - \lambda/2 $ \cite{kim2016prl}.

It is noteworthy that in the simple atomic model in Eq. \ref{AH}, we have assumed an ideal cubic crystal field with degenerate $t_{2g}$ states. In most of the real materials, however,  the TM-O$_6$ octahedral unit undergoes distortions, tilt, and rotation, removing the degeneracy of the  $t_{2g}$ states. Indeed, the resulting non-cubic crystal field is important to destabilize the atomic $J=0$ state, leading to magnetism even within an atomic model. This is straight forward to see by adding a non-cubic crystal field term, ${\mathcal H}^i_{\rm NCF}  = \sum_{lm\sigma} \varepsilon_{lm}d_{i,l\sigma}^\dagger  d_{i,m\sigma}$ in the Hamiltonian, Eq. \ref{AH}, where the onsite energies of the different orbitals are depicted by $\varepsilon_{lm}$ and the difference in the onsite energies is the measure of non-cubic crystal field $\Delta_{\rm NCF}$ . The phase diagram, Fig. \ref{fig2} (b), for the ground state shows the emergence of magnetism ($J\ne 0$) in the $d^4$ system due to  non-cubic crystal field. The competition between $\Delta_{\rm NCF}$ and SOC $\lambda$ is evident from this phase diagram.
 While in the strong SOC regimes ($\lambda \gg \Delta_{\rm NCF}$) the non-magnetic $J=0$ state remains intact, magnetism appears in the energy regime $\Delta_{\rm NCF} \gg \lambda$ [see Fig. \ref{fig2} (b)].

\subsection{Two-site model: Band structure effect} \label{model2}

\begin{figure}[t]
\centering
\includegraphics[scale=.4]{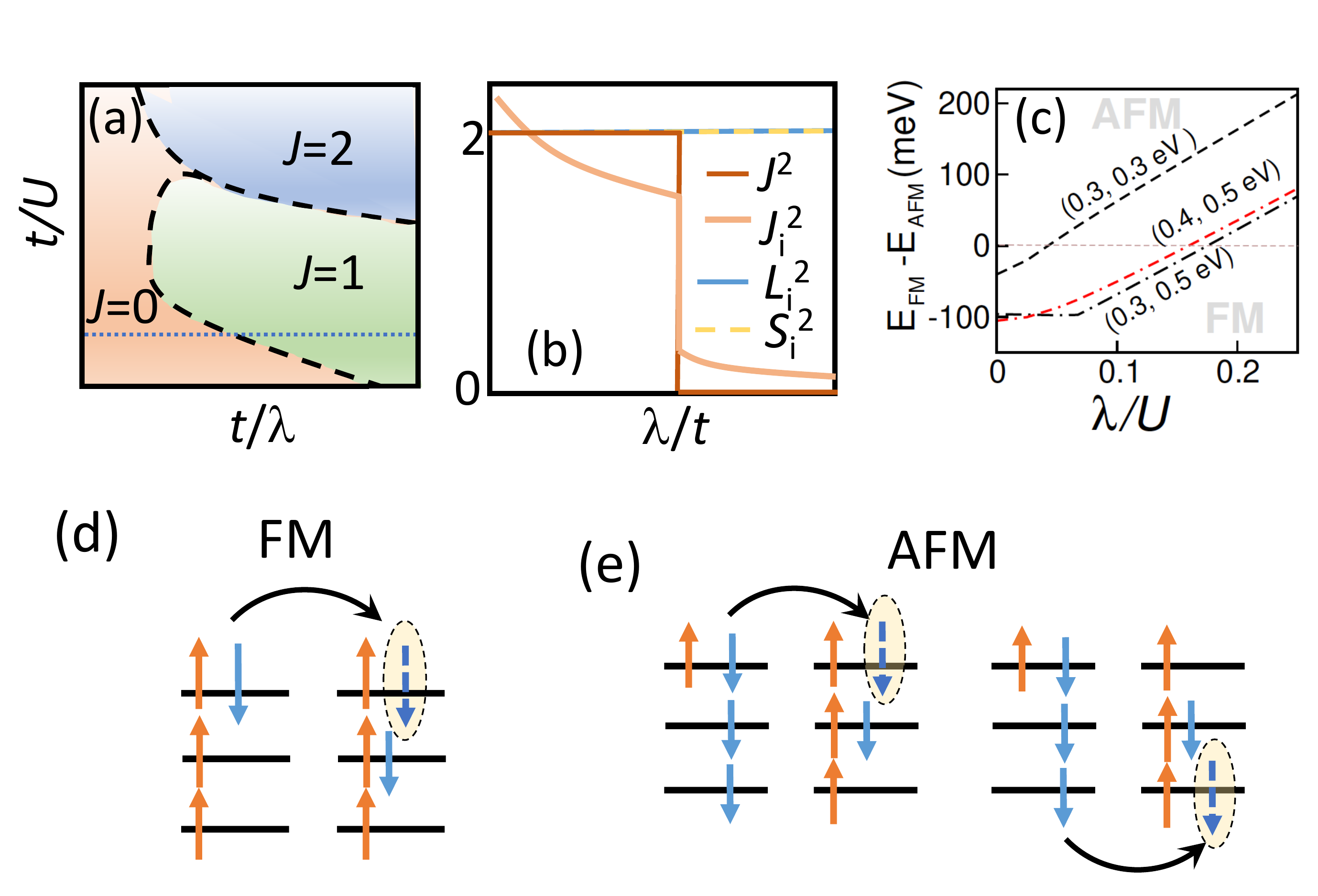}
 \caption{Schematic illustrations of the emergence of magnetism in two-site $d^4$ spin-orbit coupled model system. (a) The schematic phase diagram in the parameter space of $t/\lambda$ and $t/U$, showing the transition from the non-magnetic singlet $J=0$ state  to magnetic $J=1$ and $J=2$ states with increasing hopping. (b) The variation of the expectation values of the local $L_i^2$,  $S_i^2$ and $J_i^2$ operators and the total $J^2$ operator as functions of $\lambda/t$, indicating the discontinuous jump of the total angular momenta $J_i^2$ at each site and $J^2$ for the two-site system across the phase transition, while the local spin and orbital angular momenta $L_i^2$ and  $S_i^2$ do not change. 
 (c) The FM to AFM transition with increasing SOC for three sets of $(t,J_H)$ as indicated in the figure. 
 The different superexchange paths for the (d) FM and the (e) AFM interactions. Figures (a), (b), (d), and (e) are  reprinted with permission from Ref.  \cite{Meetei} \copyright 2015 by the
American Physical Society and (c) is reprinted from Ref. \cite{Bhowal2020}, \copyright 2020, with permission from Elsevier.
 }
\label{fig4}
\end{figure}

The atomic model, described above, misses the important ingredient of a solid which is the hopping between different atoms. The simplest model that incorporates the hopping effect is a two-site model where in addition to the atomic interactions, described in Eq. \ref{AH}, hopping between two atomic sites are also allowed. The two-site many body model Hamiltonian in a Hilbert space of $^{12}C_8 = 495$ basis states reads as 
\begin{equation}\label{two-site}
  {\mathcal H}~=~{\mathcal H}_{t} + \sum_i ({\mathcal H}^i_{int} + {\mathcal H}^i_{SO}) ,
 \end{equation}
where the first term represents the inter-site hopping between different orbitals, designated by  $t_{ij}^{l\sigma,m\sigma^\prime}$
 \begin{eqnarray}\label{hop}
 {\mathcal H}_{t} &=& \sum_{i j} \sum_{l,m} \sum_{\sigma, \sigma^\prime} 
                               ( t_{ij}^{l\sigma,m\sigma^\prime}d_{il\sigma}^\dagger d_{jm\sigma^\prime} + {\rm H.C.})
\end{eqnarray}
and the second and the third terms  in Eq. \ref{two-site} represent the atomic interactions. While the second atomic term, spin-orbit interaction, is the same as described before, the first atomic term, the interaction term, may be rewritten in terms of the total occupation number operator $\hat N_i$, total spin momentum operator $\hat S_i$, and total orbital angular momentum operator $\hat L_i$ at a 
site $i$ as  \cite{Meetei}
 \begin{eqnarray} \label{CU}
 {\mathcal H}^i_{int} &=& \frac{U-3J_H}{2} \hat N_i (\hat N_i-1 )+\frac{5}{2} J_H \hat N_i -2J_H \hat S_i^2 -\frac{1}{2} J_H \hat L_i^2.
\end{eqnarray}
A diagonal hopping in the orbital and spin space {\it i.e.}, $t_{ij}^{l\sigma,m\sigma^{\prime}}\rightarrow t\delta_{lm}\delta_{\sigma\sigma^{\prime}}$ is assumed, where all three $t_{2g}$ orbitals  actively participate in the hopping. The exact diagonalization results of the above Hamiltonian, as discussed in Ref. \cite{Meetei}, are more reliable in the Mott limit ($U/t \gg 1$). 
In this regime, a magnetic phase transition occurs from the non-magnetic $J=0$ state to a magnetic $J=1$ state. 
 The phase diagram in the parameter space of $U$, $\lambda$ and $t$ for a chosen value of  $J_H$= $0.2U$  is shown in Fig. \ref{fig4} (a).  
Interestingly, the phase transition $J=0 \rightarrow J=1$ at a critical value of  $\lambda/t$ is associated with abrupt changes in both the expectation values of individual $J_i$ and total $J$ quantum numbers across the transition.  The local $L_i$ and $S_i$ expectation values, however, remain unchanged throughout [see Fig. \ref{fig4} (b)].  In the magnetic state the local moments are produced by hopping (superexchange) driven mixing of the onsite singlet state with the higher energy triplet states. It is important to note, in contrast to standard 3$d$ Mott insulators, here the local moments are not robust and deviate from their atomic values.

The magnetic transition, discussed above for orbitally symmetric hopping, assumes full rotation symmetry which is usually absent in a real material. For example, in simple cubic systems where the hopping $t$ between TM atoms occur from the superexchange via the oxygen atoms, only two orbitals participate in the hopping  (see Fig. \ref{fig3}), so that $t_{ij}^{l\sigma,m\sigma^{\prime}}\rightarrow t\delta_{lm}(1-\delta_{km})\delta_{\sigma\sigma^{\prime}}$. Here the direction of the line, connecting the two sites $i$ and $j$, determines the blocked orbital $k$.
On the other hand, for the hopping between nearest neighbor atoms on a face-centred cubic lattice may be approximated by only one active orbital $k$, viz., $t_{ij}^{l\sigma,m\sigma^{\prime}}\rightarrow t\delta_{lm}\delta_{km}\delta_{\sigma\sigma^{\prime}}$. Here, the active orbital is determined by the common plane, shared by the two sites. 
 Interestingly, the magnetic phase transition occurs for both these cases,  emphasizing the crucial role of hopping in generating the magnetism in $d^4$ spin-orbit coupled systems \cite{Svoboda}.
A next question that naturally occurs is the nature of the magnetism.  Meetei {\it et. al} \cite{Meetei} showed that the generated local moments favor ferromagnetism 
and a phase transition from the AFM superexchange to the FM superexchange takes place at a critical value of $J_H/U$. An intuitive picture for the Hund's coupling driven ferromagnetism, may be followed from Ref. \cite{Meetei}, based on the simple perturbative analysis. The atomic ground state in the $d^4$-$d^4$ configuration with $L_i = 1$ and $S_i = 1$ has a  total energy, $E_0=12U-26J_H$, computed using Eq. \ref{CU}.  The hopping in Eq. \ref{hop} acts as a pertubation and the ground state gains energy via virtual hopping. For FM superexchange, the intermediate $d^3$-$d^5$ state [see Fig. \ref{fig4} (d)] with $S_1 = 3/2, L_1 = 0, S_2 =1/2,$ and $L_2 = 1$ has the total energy $E_{\rm FM}= 13U-29J_H$. This results in an energy gain by the FM state $\Delta E_{\rm FM} = -\frac{2t^2}{U-3J_H}$. Similarly,  the intermediate paths for the AFM superexchange, as schematically illustrated in Fig. \ref{fig4} (e), gains an energy $\Delta E_{\rm AFM} =-\frac{2}{3} \frac{t^2}{U-3J_H} -\frac{14}{6} \frac{t^2}{U}-\frac{t^2}{U+2J_H} $. Comparing the two energy scales $\Delta E_{\rm FM}$ and $\Delta E_{\rm AFM}$, it is easy to argue that the superexchange gradually changes from the AFM to FM state with increasing value of $J_H/U$. 
It is to be noted that  in the above analysis, SOC is ignored and inclusion of SOC tends to  compete against the ferromagnetism.

%
An independent exact diagonalization study \cite{Bhowal2020} with the similar two-site model including also the effect of non-cubic crystal field, also suggests the competition between ferromagnetism and  AFM with SOC in $d^4$ systems. While ferromagnetism is stabilized in absence of SOC ($\lambda =0$), with increase in $\lambda$ ferromagnetic interaction become weaker, and eventually at a critical value $\lambda_c$, an AFM interaction is favored. The critical value of $\lambda_c$ depends on the strength of the other parameters of the Hamiltonian as shown in Fig. \ref{fig4} (c). In particular, a larger Hund's coupling tends to increase the  $\lambda_c$, thereby favoring the ferromagnetism.  Dynamical mean-field theory study combined with the continuous-time quantum Monte Carlo \cite{Kim2017} also supports these findings. 

Inspired by the results of the two site exact diagonalization, the lattice problem of magnetism in $d^{4}$ systems has also been analysed by constructing an effective spin-orbital lattice model  \cite{Meetei,Svoboda,Feng}.

The effective model was employed by  Svoboda {\it et. al} \cite{Svoboda} for the analysis of magnetism for the three-different cases of active $t_{2g}$ orbitals, mentioned before,  viz., three active orbitals  ($N_{orb} =3$), two active orbitals ($N_{orb} =2$) and single active orbital ($N_{orb} =1$). 
While the detail discussion of the effective magnetic models, developed from the full Hamiltonian in Eq. \ref{two-site}, can be found in Ref. \cite{Svoboda} for each of the three mentioned cases, here we focus on the key finding of this analysis.
According to this study, while ferromagnetism is stabilized at large $J_H/U$ limit for $N_{orb} =2$ and $3$ cases, for the single active orbital case, ferromagnetic interaction is not supported for any reasonable value of $J_H/U$. The authors argue that the absence of ferromagnetism in $N_{orb} =1$ is due to the fewer available exchange paths. As illustrated in Fig. \ref{fig4}, the number of superexchange paths supporting a FM state is less than the corresponding number of such paths available for the AFM state. However, there are more possible active paths  in $N_{orb} =2$ and $3$ to reduce the energy of the FM state so that Hund’s coupling takes over, favoring ferromagnetism. This is, however, not the case for $N_{orb} =1$, where the energy of the FM state is lowered by only a single factor of $- t^2/U$, failing to favor ferromagnetism.

\begin{figure}[t]
\centering
 \includegraphics[scale=.35]{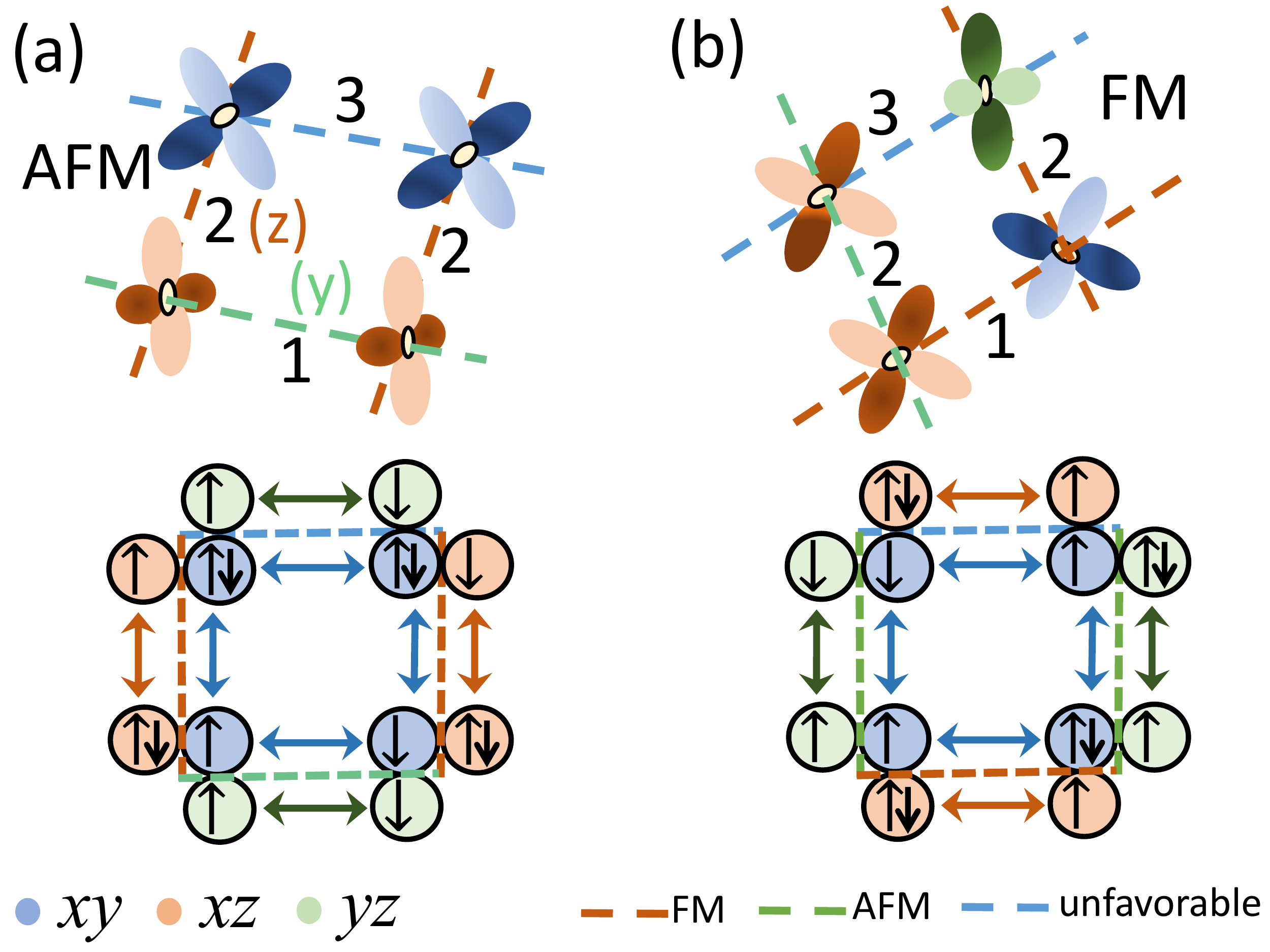}
 \caption{Graphical illustration of the orbital frustration for $N_{orb} =2$ model. The orbital frustration may be described by fixing the type of interaction along bond 1 to either (a) AFM or (b) FM. Once the interaction along bond 1 is fixed, it can be seen from (a) and (b) that the interaction along bond 3 corresponds to a higher energy configuration, leading to orbital frustration. The double occupied orbitals at each site of the square lattice are indicated in the figure. The schematic illustrations in (a) and (b) are reprinted with permission from Ref. \cite{Svoboda} \copyright 2017 by the
American Physical Society. }
\label{fig6}
\end{figure}

In addition, Ref. \cite{Svoboda} also predicts an intriguing {\it orbital frustration} for the cases of $N_{orb}=1$ and $2$ with broken rotational symmetry in contrast to the orbital symmetric $N_{orb}=3$ case.  The orbital frustration is demonstrated on a non-frustrated crystal geometry, as illustrated in Fig. \ref{fig6}, considering the case of $N_{orb}=2$. In a $d^4$ configuration, each site has at least one doubly occupied $t_{2g}$ orbital. For an AFM interaction between two sites, each with two active orbitals  ($N_{orb}=2$), the doubly occupied orbitals on both sites remain inactive [see Fig. \ref{fig4} (e)]. As a result, inter-site AFM interaction in the $N_{orb}=2$ model favors the orbital, perpendicular to the line joining these two sites, to be doubly occupied, e.g., $d_{xz}$ orbital in Fig. \ref{fig6} (a) is doubly occupied for the bond 1, along $y$-direction. These doubly occupied orbitals are, however, not inactive for bond 2,  say along $z$-direction, simply because $d_{xz}$ orbital is not perpendicular to this second bond along $z$. As a result, AFM interaction is no longer favorable and, therefore, the next energetically favorable FM interaction occurs along bond 2. This FM interaction, in turn, leads to a different doubly occupied orbital ($d_{xy}$ orbital in Fig. \ref{fig6}) for  the bond 3 along $y$-direction, which neither favors energetically lowest FM interaction nor the energetically lowest AFM interaction, stabilizing rather an energetically higher AFM interaction along bond 3. Similarly, starting from the FM interaction between two sites along a particular bond [see bond 1 in Fig. \ref{fig6} (b)], it is possible to show that a higher energy configuration is taken by one of the four bonds on a plaquette.  This leads to a frustration due to orbital degrees of freedom even on a non-frustrated lattice, which may result in an orbital liquid state in the regime $\lambda \ll zJ_{SE}$ ($z$ being the coordination number) with strong orbital effects  in absence of strong octahedral distortions.

\begin{figure}[t]
\centering
\includegraphics[scale=.4]{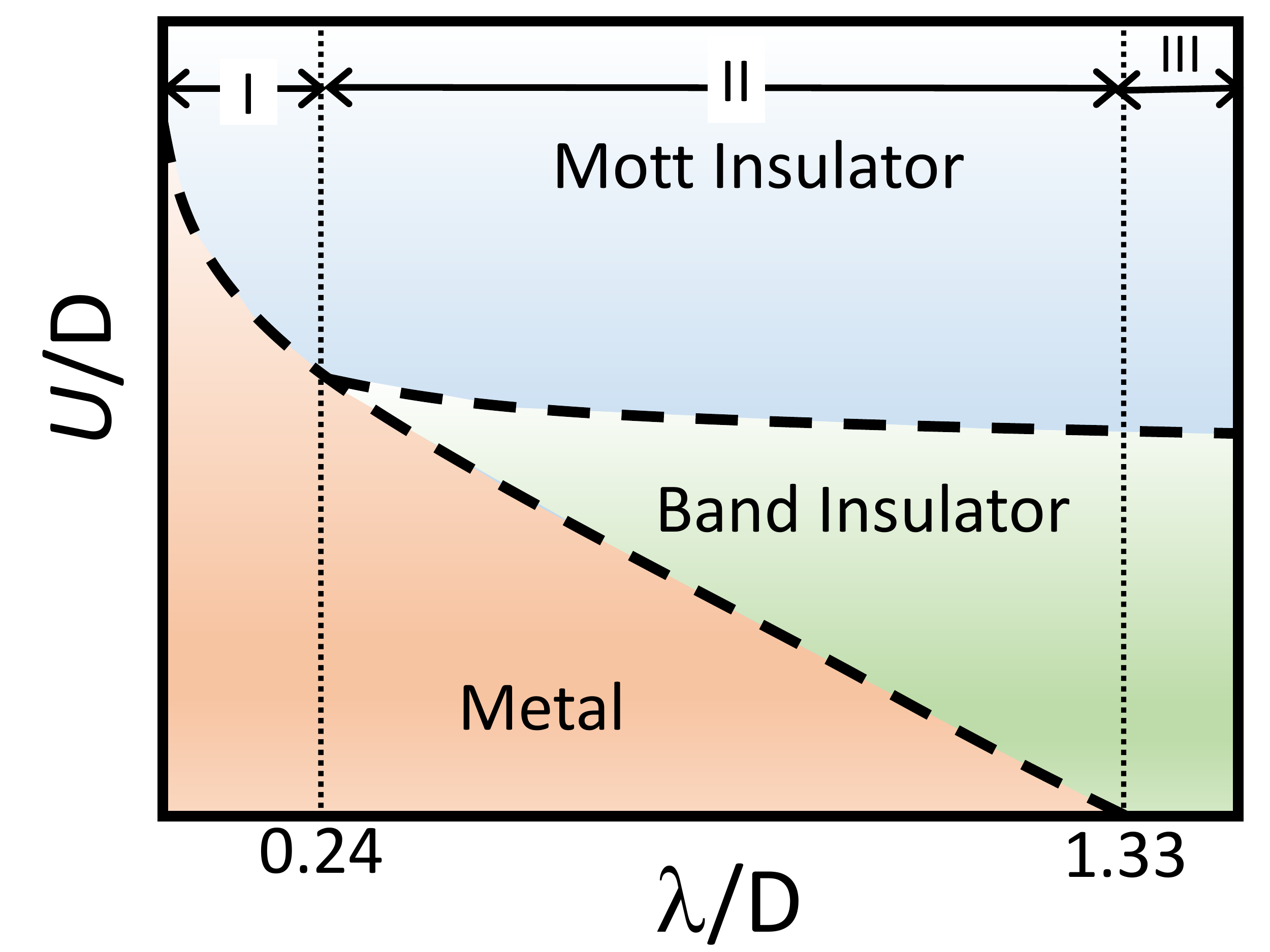}
 \caption{Schematic illustration of the phase diagram, adapted with permission from Ref. \cite{Du2013} ~\copyright 2013 Springer Nature, showing the metal to insulator transition in the parameter space of $U/D$ and $\lambda/D$, where $D$ is the half-bandwidth.}
\label{fig7}
\end{figure}

\subsection{Metal-insulator transition}
We now turn to the transport properties of $d^4$ spin-orbit coupled systems. 
The transport properties in $d^4$ systems, as studied in Ref. \cite{Du2013} using 
 the rotational invariant Gutzwiller approximation (RIGA) method as well as the dynamical mean-field theory (DMFT). The calculations reveal  metal-insulator transition as a function of SOC constant $\lambda$ and Coulomb interaction $U$.
 A Hubbard model for the $t_{2g}$ orbitals with full Hund’s rule coupling and spin-orbit interaction, similar to the many-body model, discussed above, is considered in this work. The computed phase-diagram in the parameter space of  $\lambda$ and  $U$ (see Fig. \ref{fig7}) depicts the three different states, viz., metallic state, Mott insulating state, and band insulating state. The two insulating states can be visualized by considering the two limiting cases. First, let us consider the strong SOC limit ($U=0$). In this limit, the degenerate $t_{2g}$ bands split into $j_{\rm eff} =3/2$ and $j_{\rm eff} =1/2$ bands, separated from each other with an energy gap of $\approx 3\lambda/2$. With increase in $\lambda$, the higher lying $j_{\rm eff} =1/2$ bands tend to get empty as the electrons move to the  $j_{\rm eff} =3/2$ bands. Consequently, at a critical value of $\lambda/D$, $D$ being the half-bandwidth, the  $j_{\rm eff} =3/2$ bands are completely occupied, resulting in a transition to the band insulating state from a metallic state. Moving to the other limit where $\lambda=0$, the degenerate $t_{2g}$ orbitals have four electrons at each site. With increasing $U$, the system undergoes an interaction driven  Mott transition at some critical value of $U/D$ with each band having $4/3$ electrons.  

Coming back to the discussion of the phase diagram in Fig. \ref{fig7}, we can see that the phase diagram may be subdivided into three different regions: (I) The regime $0<\lambda/D<0.24$, where increase in $U$ leads to a transition from metal to Mott insulator. The decrease in the critical value of $U$ for this metal-insulator transition with increasing SOC, suggests the enhancement of the correlation effects by the SOC. Intuitively, this may be attributed to the narrow bands, resulting from the SOC driven band splitting. It is interesting to point out that similar cooperation between SOC and correlation has been observed for $d^5$ filling as well \cite{Liu2008}. In this Mott phase, the local moment at each site vanishes, forming a spin-orbital singlet state and the ground state is simply the product state of the local singlets at each site. (II) The next regime is  $0.24<\lambda/D<1.33$. Two back to back transitions occur in this regime as we increase $U$. The first transition 
is from metal to band insulator, followed by the second transition from band to Mott insulator. For  the intermediate $U$, the effective bandwidth of the system is reduced by the correlation effect first, which upon relatively small band splitting due to intermediate $\lambda$ results in a band insulating state. Increasing $U$ further pushes the system into Mott regime. (III) Finally, in the regime   $\lambda/D >1.33$, strong SOC leads to completely empty $j_{\rm eff} =1/2$ bands even at $U=0$, suggesting a band insulating state in the non-interacting case. Increasing $U$ to a large value induces again a transition from band insulating state to Mott insulating state. 

\section{Materials}

The various theoretical predictions \cite{Khaliullin, Meetei} described in the previous section, for the $d^4$ spin-orbit coupled systems generated enormous interest in pentavalent iridates for the realization of excitonic magnetism. The search  is further, triggered by the observation of novel magnetism in $d^4$ double perovsite system Sr$_2$YIrO$_6$ \cite{CaoSYIO},  followed by a series of other $d^4$ iridates. The observed magnetism in iridates is, however, quite enigmatic. The ground state properties of these materials are full of controversial and conflicting results. While a large number of materials have been studied and characterized in past years, here we focus on a few particular materials that have received considerable attention in the $d^4$ iridate research.     

The pentavalent iridates, that we discuss in this review, can be broadly categorized into four classes according to their crystal geometry : (a) double perovskite (b) post perovskite (c) 6H perovskite and (d) hexagonal iridates. 
The crystal structure plays a key role in deciding  magnetism in iridates, so the ground state properties of the iridates, belonging to the same geometry class, are expected to be quite similar. Interestingly, the ground state of $d^4$ iridates is so delicate that even small changes in the fine details of the structure can lead to dramatic changes in the ground state properties, 
as we now proceed to discuss below.

\subsection{Double perovskite iridates} \label{DP}

\begin{figure}[t]
\centering
 \includegraphics[width=\linewidth]{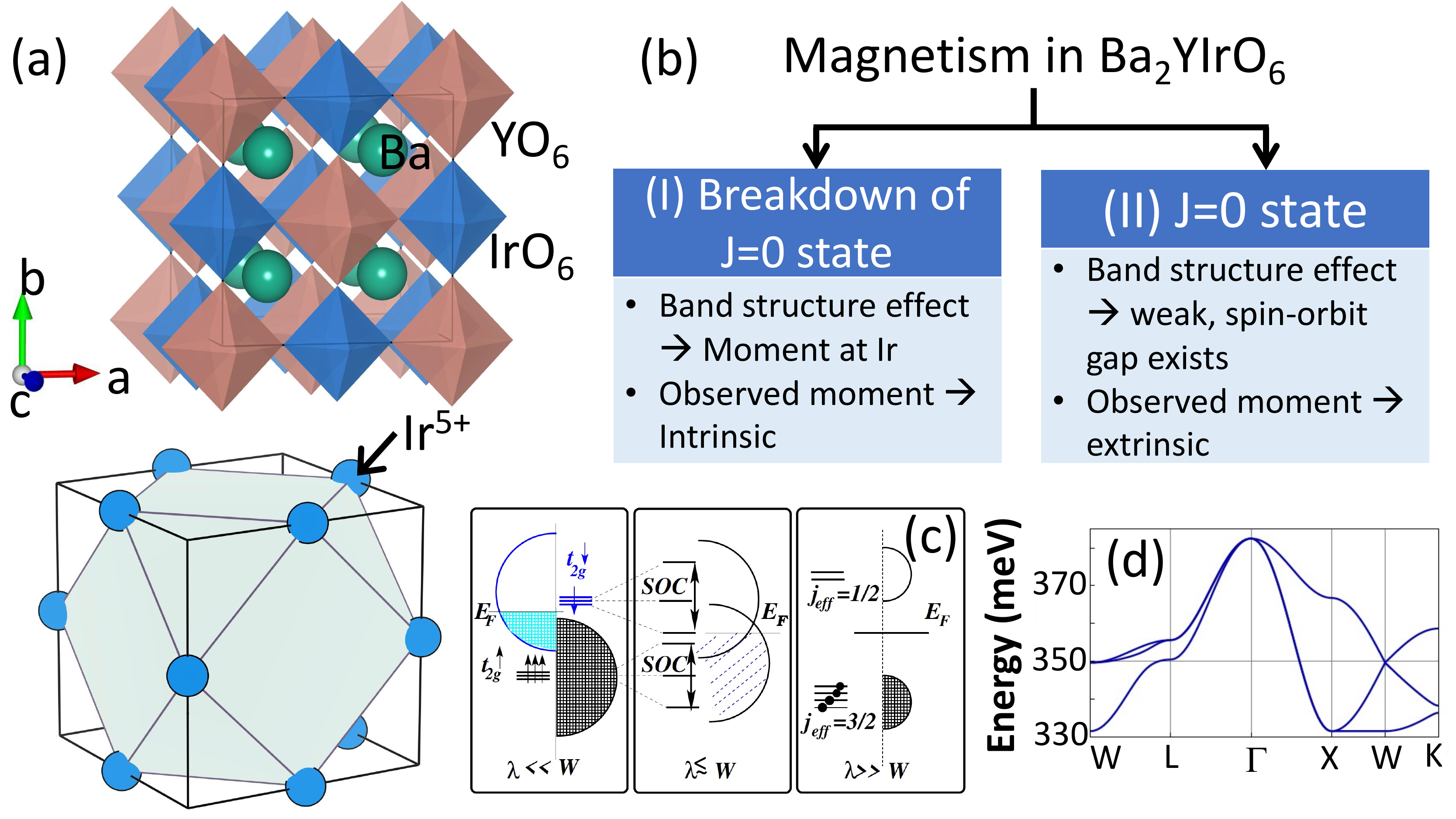}
 \caption{Schematic illustration of the two different scenario, proposed for double perovskite iridate Ba$_2$YIrO$_6$. (a) Crystal structure of a double perovskite iridate ({\it top}) and the geometric frustration of the face-centered lattice formed by the Ir ions ({\it bottom}). (b) The summary of the two conflicting scenario, discussed in theory and experiment, regarding the magnetism in Ba$_2$YIrO$_6$. While the scenario (I) supports the breakdown of the $J=0$ state, scenario (II) suggests the presence of $J=0$ state. According to (I), the experimentally observed moment is intrinsic and it appears due to band structure effect, illustrated in (c), where the double perovskite iridate lies in the intermediate regime of 
 $\lambda \le W$. The scenario (II), on the contrary, suggests absence of any moment at the Ir site 
  and the observed moment in the experiment is ascribed to the extrinsic effects, such as anti-site disorder, impurity, magnetic defect etc. The figure (d) in this scenario shows the weak dispersion of the triplet excitation spectrum, failing to overcome singlet-triplet energy gap. The figures for the first scenario is taken from Ref. \cite{Bhowal2015} and the second scenario is reprinted with permission from Ref. \cite{Chen} \copyright 2017 by the
American Physical Society. }
\label{fig8}
\end{figure}

The general formula for the double perovskite iridates, that we discuss here, is A$_2$BIrO$_6$, where the A site is occupied by the divalent cations, e.g., Sr, Ba. In this case, the trivalent ions such as Y$^{3+}$, Sc$^{3+}$, Lu$^{3+}$, La$^{3+}$, etc., at the B site renders the desired $5+$ charge state at the Ir site. 
 The basic structure of the double perovskite iridates consists of corner shared BO$_6$ and IrO$_6$ octahedra that alternate  along each of the crystallographic axes. 
 These octahedra are either ideal or undergo specific types of distortions, depending on the particular crystal symmetry of the structure.       
 In general, the crystal structure accompanies strong geometric frustration due to face-centered cubic lattice formed by the neighboring Ir atoms (see Fig. \ref{fig8}). 
 
Owing to the compositional and structural flexibility, double perovskite iridates have particularly received  enormous attention. The interest in  the $d^4$ double perovskite iridates began with the observation of the onset of magnetism in double perovskite iridate Sr$_2$YIrO$_6$ in the year 2014, although, the magnetism in iridates with $d^4$ electronic configuration was reported even earlier, e.g.,  in another double perovskite system Sr$_2$CoIrO$_6$ \cite{Narayanan} (this is even before the theoretical prediction by Khaliullin \cite{Khaliullin}). In this work  Narayanan {\it et. al.} \cite{Narayanan} studied the structural, electronic and magnetic properties of La$_{2-x}$Sr$_{x}$CoIrO$_6$ for $0 \le x \le 2$, the end member of which is Sr$_2$CoIrO$_6$ with high spin Co$^{3+}$ and low spin Ir$^{5+}$ ions. Sr$_2$CoIrO$_6$ undergoes successive structural transitions starting from a two-phase region upon heating 
, viz., $P2_1/n+I2/m  \rightarrow I2/m \rightarrow I4/m \rightarrow Fm\bar{3}m$, and exhibits 
AFM ordering as revealed from magnetization measurements.
Interestingly, the estimated effective magnetic moment for the system $\mu_{\rm eff} = 5.1 (1) \mu_B/$ f.u. suggests for magnetic contribution from the Ir$^{5+}$ ion, as the contribution ($\mu_{\rm eff} = 4.9 \mu_B/$ Co) for the Co$^{3+}$ ion alone can not describe the estimated moment value. This small moment, however, remains undetected in the neutron powder diffraction (NPD) measurements.  Nevertheless, the density functional calculations, presented in the same work, also suggest significant spin and orbital moments at the Ir site. %
However, the material is reported to have significant anti-site disorder and, also, the presence of strongly magnetic Co$^{3+}$ ions brings up the question if the Co ions have any role in the magnetism at the Ir site. Later, it was also pointed out that in addition to Ir$^{5+}$ ion, the system also has Ir$^{6+}$ ions, which have rather strong magnetic signal \cite{stefano}.  In view of the above, the  double perovskite iridates where the magnetism can appear only via Ir (B site is non-magnetic) are the best candidates for the investigation of magnetism in the Ir$^{5+}$ ion.

\begin{figure}[t]
\centering
\includegraphics[scale=.6]{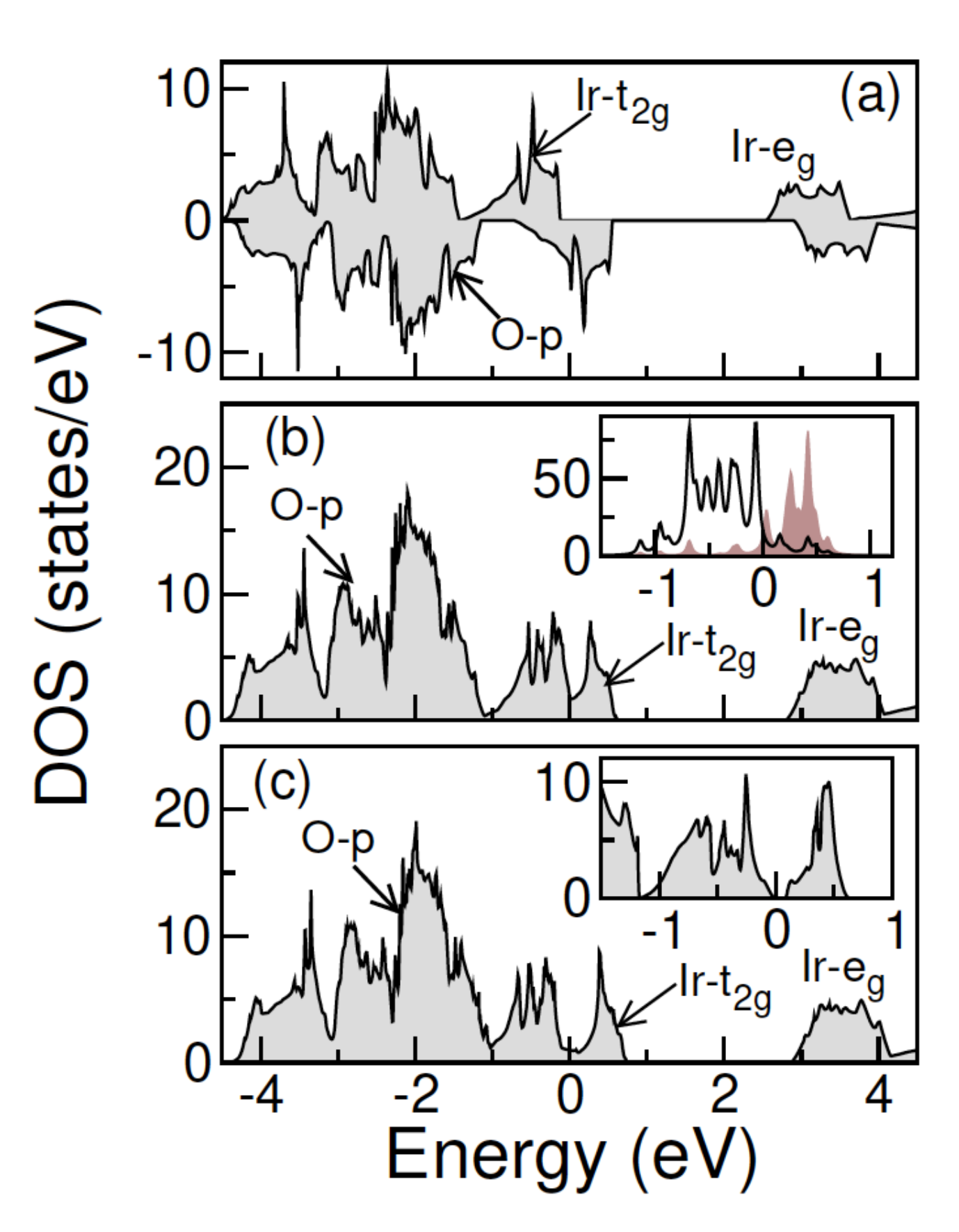}
 \caption{The densities of states (DOS) for Ba$_2$YIrO$_6$, computed within (a) GGA, (b) GGA+SOC, and (c) GGA+SOC+U, considering ferromagnetic spin configuration. The dominant orbital characters are indicated in the figure. The GGA spin polarized calculation gives a half-metallic solution in (a). Inclusion of SOC leads to changes in the DOS, as indicated in (b). The inset in (b) shows the mixing between $j_{\rm eff} =1/2$ (shaded region) and $j_{\rm eff} =3/2$ (solid line) states, obtained from a non-magnetic Ir-$t_{2g}$ only tight-binding model in presence of SOC, indicating the origin of finite moment at the Ir-site. Application of Coulomb interaction $U$ fails to open up a gap within the ferromagnetic spin configuration, while the energy gap is present in the computed energetically lowest type-I anti-ferromagnetic configuration as shown in the inset of (c). }
\label{figdos}
\end{figure}

A series of such double perovskite iridates A$_2$R$^{3+}$Ir$^{5+}$O$_6$  (A= Sr, Ba and R=La, Y, Sc, Lu) were synthesized long ago in 1999 by Wakeshima {\it et. al.} \cite{Wakeshima}  and all of these compounds were reported to have paramagnetic behavior down to 4.5 K. With the new insight into the SOC effect in iridates, later, Cao {\it et. al.} \cite{CaoSYIO} revisited the compound Sr$_2$YIrO$_6$  and showed that the system undergoes a novel magnetic transition below $T=1.3$ K. The observed magnetism was surprising as both the presence of Ir$^{5+}$ ion as well as strong geometric frustration in the double perovskite structure in general should forbid any long range magnetic order.
The reported moment $\mu_{\rm eff}$ at the Ir site is  0.91 $\mu_B$ and the Curie-Weiss temperature $\theta_{\rm CW}$ is as large as $-229$ K, indicating the presence of AFM interaction as well as strong frustration in the system. Sr$_2$YIrO$_6$ crystallizes in monoclinic $P2_1/n$ symmetry, where the IrO$_6$ octahedron  undergo distortion. 
 The authors argued the non-cubic crystal field, resulting from this distortion, as the possible origin of deviation from the singlet $J=0$ state, leading to magnetism in Sr$_2$YIrO$_6$.
 
 Interestingly,
the parallel study of another $d^4$ double perovskite iridate Sr$_2$GdIrO$_6$ in the same work \cite{CaoSYIO} showed that
unlike the Y-compound, the Gd-compound orders ferro-magnetically at $T_c = 2.3$ K with a
comparable Curie Weiss temperature $\theta_{\rm CW} = 1.2$  K. The saturation moment was found
to be $M_s \sim 7 \mu_B$/fu consistent with the Gd$^{3+}$ (4$f^7$) charge state, present in the
system. The authors proposed Ir atoms in Sr$_2$GdIrO$_6$ to be non-magnetic, however the presence of a strong magnetic signal from Gd$^{3+}$ ion makes the confirmation difficult. According
to this study, the latter iridate crystallizes in cubic space group $Pn$-$3$
and, therefore, is free from any non-cubic distortion. In absence
of any non-cubic crystal field, SOC dominates, resulting in a $J=0$ state at the Ir site in Sr$_2$GdIrO$_6$.
 Both double perovskite iridates were reported to be insulating with an
energy gap of 490 meV and 272 meV respectively.

The magnetic properties of these two double-perovskite iridates were later studied within the framework of  density functional theory \cite{Bhowal2015}. The electronic structure calculations presented in this work reveal that in addition to  non-cubic crystal field the band structure effects are,  more crucial for the breakdown of the anticipated $J=0$ state leading to the formation of moments. It is interesting to point out that the crucial role of the band structure effects  has also been discussed in the context of tetravalent  iridates \cite{Zhangprl2013}.
The conclusions of Ref. \cite{Bhowal2015}  is based on the estimated non-cubic crystal field and the magnetic moments at the Ir site in the two double perovskite iridates, mentioned above,  along with another double perovskite, Ba analogue of Sr$_2$YIrO$_6$, Ba$_2$YIrO$_6$. The authors showed that even though Sr$_2$GdIrO$_6$ crystallizes in the cubic structure, the $Pn-3$ symmetry
allows for the rotation of the IrO$_6$ octahedra, that leads to a strong trigonal splitting of the Ir-$t_{2g}$ orbitals, which is about an order of magnitude larger than that in  Sr$_2$YIrO$_6$. 
 Indeed, the electronic structure
calculation showed the presence of finite moment at the Ir site in both these systems. The
situation became even more interesting with the cubic double perovskite iridate
Ba$_2$YIrO$_6$, where the $Fm\bar{3}m$ symmetry does not allow any tilt or rotation of the IrO$_6$ octahedra, resulting in a triply degenerate $t_{2g}$ orbitals. Even in this case, finite moment at the Ir site persists. This emphasizes the band structure effects to be crucial in driving magnetism in these series of double perovskite iridates [see scenario I in Fig. \ref{fig8} (b)]. Interestingly, despite the different non-cubic crystal field, the bandwidths in this series of double perovskites are comparable and large in magnitude ($\sim 1.2$ eV)  that leads to significant mixing between $j_{\rm eff} =1/2$ and $j_{\rm eff} =3/2$ states [see inset of Fig. \ref{figdos} (b)], resulting in  deviation from the ideal atomic limit. The details of the evolution of the electronic structure of Ba$_2$YIrO$_6$, obtained in this work, is summarized in Fig. \ref{figdos}.

In complete contrast to  Refs. \cite{CaoSYIO,Bhowal2015}, DMFT calculations \cite{Kunes} suggest absence of any long range magnetic order in both the compounds Sr$_2$YIrO$_6$ and Ba$_2$YIrO$_6$.   
 In the above 
work, taking density functional theory as a starting point, within the constrained
random phase approximation, a multi band Hubbard model is constructed which is
further analyzed within static and dynamic mean field theories and strong coupling
expansion. The authors showed that a large $U$ ($U=4$ eV) is required to get an insulating solution, reducing significantly the mixing between $j_{\rm eff} =1/2$ and $j_{\rm eff} =3/2$ states that was present within GGA+SOC. Using effective strong coupling model, the authors concluded that the reduced magnitude of hopping due to large Ir-Ir distance in the double perovskite structure is too small to overcome the singlet-triplet energy gap, which was predicted to be in the range of 200 meV.  This results in a non-magnetic $J=0$ singlet ground state in both the iridates. 

Around the same time, Dey {\it et. al.} \cite{TDey} revisited and studied in detail the structural, magnetic, and thermodynamic properties of Ba$_2$YIrO$_6$, which was originally synthesized more than 45 years ago \cite{Thumm}. This study confirmed the ideal cubic structure of  Ba$_2$YIrO$_6$, in contrast to the monoclinic symmetry reported earlier \cite{Wakeshima}, and showed that the system does not order down to 0.4 K. However, the Curie-Weiss fit yields a Curie Weiss temperature $\theta_{CW} \sim -8.9$ K and an effective moment of $0.44 \mu_B$/Ir, unexpected for a $J = 0$ state.  The authors exclude the possibility of any anti-site disorder due to different sizes of Ir$^{5+}$ and Y$^{3+}$ ions or oxygen vacancy from their magnetization measurements on the oxygen-annealed crystals. On the contrary, the DFT calculations presented in the same  work show a non-magnetic ground state in Ba$_2$YIrO$_6$ with significant mixing between $j_{3/2}$ and $j_{5/2}$ states (the effective $j_{\rm eff} =1/2$ state is part of this full $j_{5/2}$ manifold \cite{Ganguly}).  While in absence of Hubbard $U$, the system remains metallic, the insulating state is obtained with a rather small value of $U$, viz., $U = 1.4$ eV and $J_{\rm H}= 0.5$ eV in contrast to Ref. \cite{Kunes}. The origin of the moment at the Ir site remains elusive from this work. 

In the meantime, independent experiments \cite{Ranjbar,Phelan} probed the effect of structural distortions on the magnetism of the series of double perovskites Ba$_{2-x}$Sr$_x$YIrO$_6$ ($0 \le x \le 1$). Both studies show no evidence for magnetism down to 2 K. The effective moment, reported in Ref. \cite{Ranjbar}, is rather small, 0.16 $\mu_B$/Ir and was speculated as a result of local disorder in the structure which was not captured in the diffraction measurements. These results were, further, disputed by Terzic {\it et. al.} \cite{Terzic}, emphasizing that the realization of the magnetic ground state requires investigation below 2 K. They confirmed the presence of magnetic order in the single-crystal, double-perovskite Ba$_2$YIrO$_6$ and Sr-doped Ba$_2$YIrO$_6$ below 1.7 K using dc magnetization, ac magnetic susceptibility, as well as  heat-capacity measurements. 
From a  fitting over a wide temperature range $50 < T <300$ K in this work, the authors estimated an effective moment of 1.44 $\mu_B$/Ir in Ba$_2$YIrO$_6$ and a large Curie-Weiss temperature  of -149 K.  The authors questioned the large singlet-triplet gap predicted earlier \cite{Kunes}, claiming that such a large gap does not explain the significant temperature dependence of the magnetic susceptibility at low temperature. 
 
The debate continued as, in the same year, Corredor {\it et. al.} \cite{Corredor} even questioned the previously reported monoclinic symmetry of Sr$_2$YIrO$_6$ \cite{CaoSYIO,Ranjbar} and proposed instead the same cubic structure as its Ba analogue.  According to this work Sr$_2$YIrO$_6$ does not undergo any magnetic transition down to 430 mK. They pointed out the low-temperature anomaly in the specific heat as a Schottky anomaly due to paramagnetic impurities, and is, therefore, not  related to the onset of magnetic order. The small effective magnetic moment $\mu_{\rm eff} = 0.21 \mu_B$/Ir and the small correlation were attributed to the $J =1/2$ impurities in the sample. Along the same line,  the observed moment in Ba$_2$YIrO$_6$ was ascribed to extrinsic origin, such as paramagnetic impurity \cite{Hammerath}, anti-site disorder \cite{Chen}, Ir$^{4+}$/Ir$^{6+}$ magnetic defects \cite{Fuchs}, excluding the possibility of any excitonic magnetism [see scenario II in Fig. \ref{fig8} (b)].

On the contrary, Nag {\it et. al.} \cite{Nag2018} pointed out that the temperature dependence of the magnetic response below 10K for Ba$_2$YIrO$_6$ as revealed from their $\mu$SR studies is rather  unusual. 
In particular, the  temperature dependence of the stretched exponent parameter $\beta$ of the fitted  muon polarization P(t) = $ \exp-\left( \alpha t\right)^{\beta}$ cannot be corroborated with  a $J=0$ state, for which no such temperature dependence of $\beta$ should exist. The observed behavior is explained as follows:
while a small iridium moment exists ubiquitously in Ba$_2$YIrO$_6$, at higher
temperature the system behaves as a paramagnet. When the temperature is lowered below 10 K,  
most of these Ir moments start to interact with each other anti-ferromagnetically and pair up to form spin-orbital singlets, leaving behind a few Ir ions due to Y/Ir disorder and with a very low residual interaction between them. This is in sharp contrast with the previous proposal where the impurity in the $J=0$ background has been suggested for Ba$_2$YIrO$_6$. 
The proposed scenario of Nag  {\it et. al.} is quite similar to its 4$d$ analogue Ba$_2$YMoO$_6$ \cite{Vries}, supporting the scenario I in Fig. \ref{fig8} (b)].

Clearly the ground states of  Sr$_2$YIrO$_6$ and Ba$_2$YIrO$_6$ are quite controversial. The situation becomes even more puzzling with the report of resonant inelastic x-ray scattering (RIXS) measurements in these double perovskite systems \cite{Nag2018,Yuan} that facilitate a quantitative estimation of the parameters $\lambda$ and $J_H$ in these systems. 
We shall discuss the RIXS technique in some detail later in the context of 6H perovskite iridates.
Interestingly, regardless of the presence of the non-cubic crystal field (Sr vs Ba compound) or its strength (Y vs Gd compound), the RIXS spectra in all these double perovskite iridates are quite similar and can be described by an atomic  model with $\lambda \approx 0.39-0.42$ and $J_H \approx 0.24-0.26$ eV \cite{Nag2018,Yuan}. While theoretical studies \cite{Chen,Gong2018,BHKim} suggest excitonic magnetism is unlikely in Ba$_2$YIrO$_6$ due to large estimated singlet-triplet gap in RIXS measurement, Ref. \cite{Terzic} indicates 
the comparable energy scales of Hund's coupling and SOC, as a possible factor to drive  magnetism in these double perovskite iridates. Furthermore, recent DFT calculations \cite{Paramekanti}  show the presence of type-I AFM state in Ba$_2$YIrO$_6$, similar to the initial DFT results \cite{Bhowal2015}, confirming the possibility of interesting physics beyond the atomic limit. The insulating state in Ref. \cite{Paramekanti}, however, occurs at a higher $U$ than in Ref. \cite{Bhowal2015} ($U = 4$ vs $U = 2$ eV). The magnetism in Ref. \cite{Paramekanti} is attributed to the “intracluster excitons” that arise from the Ir-O interaction within an IrO$_6$ cluster. 

The formation of moments at the iridium site was also discussed in the Sc counterparts of Sr$_2$YIrO$_6$ and Ba$_2$YIrO$_6$, viz.,  Sr$_2$ScIrO$_6$ and Ba$_2$ScIrO$_6$~\cite{Kayser,Chakraborty2019,Bhowal2020}. Similar to the Y compounds,  Ba-Sc compound crystallizes in the cubic $Fm\bar{3}m$ structure while Sr-Sc compound undergoes monoclinic distortion of the $P2_1/n$ space group. Experimentally, the systems do not exhibit long range magnetic order while an effective moment of 0.16 $\mu_B$/Ir for the Sr compound and even larger moment of 0.39 $\mu_B$/Ir for the Ba compound was estimated.  The generated magnetic moment was attributed to the hopping induced de-localization of holes in these systems \cite{Chakraborty2019,Bhowal2020}. 

The compositional flexibility within the general formula of A$_2$BIrO$_6$ leads to a large number of pentavalent double perovskite iridates, that includes but not limited to the listed materials in Table \ref{tab1}. Further flexibility is achieved by doping at the A-site or/and using a magnetic atom at the B site. While doping offers the possibility to explore the effect of the resulting structural distortion on the magnetic properties of the system, magnetic B-site offers the direct study of 3$d$-5$d$ interaction, combining the strong correlation of 3$d$ TM ion to the strong SOC of 5$d$ TM ion.
 While this offers a broad area of research spanned by some excellent works \cite{Laguna-Marco,Kharkwal,Kolchinskaya,stefano}, we do not discuss them in the present review.      

\subsection{Post-perovskite iridate}

\begin{figure}[t]
\centering
 \includegraphics[width=\linewidth]{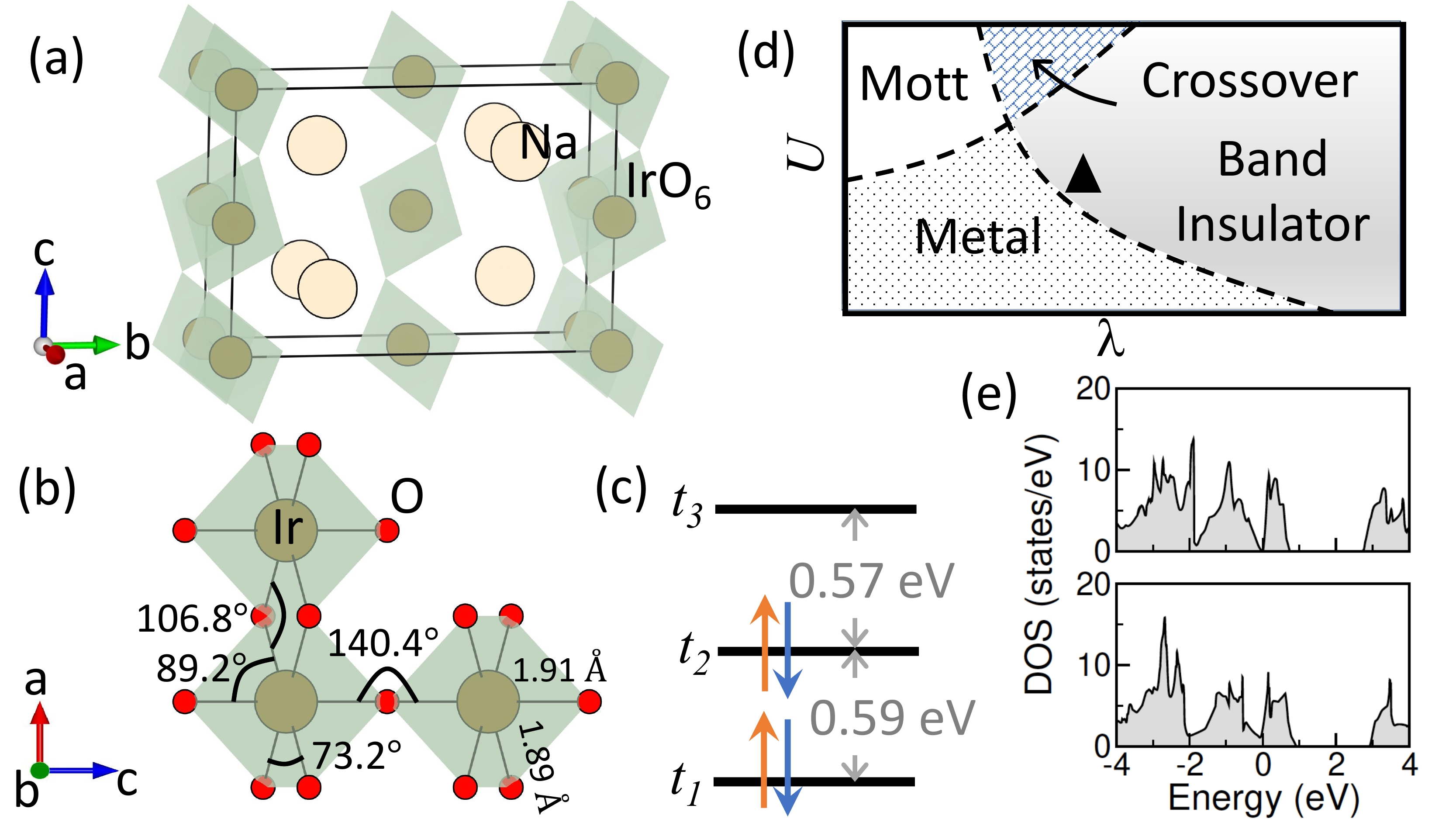}
 \caption{The $S=0$, insulating state of NaIrO$_3$. (a) The crystal structure of NaIrO$_3$ in the orthorhombic $Cmcm$ space group. (b) The distorted IrO$_6$ octahedra and their connection in the $ac$ plane. Different bond lengths and bond angles are also indicated in the figure.  (c) The resulting non-degenerate Ir-$t_{2g}$ states, $t_1, t_2$, and $t_3$, and the non-cubic crystal field splitting between the states. The large $\Delta_{\rm NCF}$ leads to $S=0$ state, rather than $J=0$ state. (d) The schematic illustration of the phase diagram, adapted with permission from Ref. \cite{Du}, in the parameter space of $U$ and $\lambda$ relevant for  NaIrO$_3$. The realistic parameters for NaIrO$_3$ ($U=2$, $\lambda=0.33$ eV, and $J_H=U/4$) correspond to a band insulating state, indicated by the black triangle in the figure. (f) The computed density of states for NaIrO$_3$, adopted from Ref. \cite{Bhowal2015}, considering the
experimentally reported crystal structure \cite{Bremholm} ({\it bottom}) and the optimized structure as in Table. \ref{tab2} ({\it top}), indicating correlation effects may not be important for the experimentally observed insulating behavior. }
\label{fig9}
\end{figure}

The first post-perovskite oxide with pentavalent atom NaIrO$_3$, containing Ir in
the $d^4 $ configuration, was synthesized using high-pressure solid state method \cite{Bremholm}. The post perovskite NaIrO$_3$ crystallizes in the layered orthorhombic $Cmcm$ space group. In contrast to the corner shared IrO$_6$ octahedra in the double perovksite iridates, discussed above, here  the structure consists 
of distorted IrO$_6$ octahedra which not only share their corners along the $c$-axis but also share the edges
along the $a$-axis and, thereby, form IrO$_3$ sheets. As shown in Fig. \ref{fig9} (a) and (b), these sheets are, further, stacked along the $b$ direction, separated by the
Na atoms.

Interestingly, the magnetic measurement as well as DFT calculations \cite{Bremholm} confirm the nonmagnetic state of Ir in NaIrO$_3$. This
was, thus, considered an ideal  example for the  realization of the $J = 0$ nonmagnetic state \cite{Bremholm}, in marked contrast with the series of compounds, discussed in the previous section.  
The electrical resistivity in this work showed that the system is non-metallic, however the DFT calculation fails to describe the insulating state observed in experiments, leaving the 
origin of the resistivity unclear. A variable
range hopping model was used to describe the low temperature resistivity of NaIrO$_3$. 
This put forward two important questions for this post-perovskite iridate: Firstly, whether NaIrO$_3$ is  a true realization of the $J=0$ ground state? Second, whether the material is closer to a “renormalized band insulator”, where the correlation reduces the effective bandwidth which upon SOC becomes gapped out, or  is it closer to “Mott insulator”, driven by strong Coulomb interaction. The DFT calculations in Refs. \cite{Bhowal2015} and \cite{Du} addressed these issues,  providing insights to the answer to these questions, which we now proceed to discuss.

\subsubsection{$S=0$ instead of $J=0$ state in ${\rm NaIrO_3}$:}

According to the theory that supports the magnetism at the Ir site in the pentavalent double perovskite iridates, as discussed in section \ref{DP}, emphasizes the crucial role of band structure effects in preventing the realization of the desired $J=0$ state. The band structure effects become particularly important due to the three-dimensional connectivity of the Ir atoms. Therefore, the reduced dimensionality of the connecting Ir atoms, resulting from the two-dimensional Ir sheets in the post perovskite structure (see Fig. \ref{fig9}) raises the hope to find the $J=0$ state in NaIrO$_3$.
 
However, in addition to the desired layered structure, the IrO$_6$ octahedra in NaIrO$_3$ also suffer from strong distortions [see Fig. \ref{fig9} (b)]. 
The $\angle $O-Ir-O bond angle deviates significantly from the ideal 90$^\circ$ angle to about $\approx$ 73.2$^\circ$, while the edge and corner sharing $\angle $Ir-O-Ir angles are respectively 106.8$^\circ$ and 140.4$^\circ$. The Ir-O bond lengths also differ by about $0.02$~\AA. Such a high  distortion leads to strong non-cubic crystal field, that removes the degeneracy of the $t_{2g}$ orbitals completely, splitting it into three non-degenerate energy levels. The energy separation between these levels were estimated to be $\Delta_{\rm NCF} \approx$ 0.57-0.59 eV \cite{Bhowal2015}, which is comparable to the typical energy scale of the Hund's coupling $J_H \approx 0.4-0.6$ eV in 5$d$ TM oxides. As a result, $\Delta_{\rm NCF}$ dominates over $J_H$, stabilizing an electronic configuration $(\uparrow\downarrow,\uparrow\downarrow,0)$ [see Fig. \ref{fig9}(c)] that leads to a non-magnetic state.    

 Clearly, the non-magnetic state in NaIrO$_3$ is distortion driven and has noting to do with the SOC. Thus, it is better described by an $S=0$ state, rather than a $J=0$ state. The stabilization of the non-magnetic state even without SOC in the DFT calculations \cite{Bremholm,Bhowal2015} further supports this finding. 
 
\subsubsection{Insulating state:} 

In contrast to the observed insulating behavior in the experiments, DFT calculations show a metallic state in the non-magnetic ground state \cite{Bhowal2015,Bremholm}. Application of SOC and Coulomb correlation $U$ do not affect this scenario. This leads to the puzzling insulating behavior in NaIrO$_3$.    

Failing to describe the insulating behavior within DFT+SOC+U approach, Ref. \cite{Du} made an attempt to investigate the insulating state with the LDA+Gutzwiller method. Starting from a low energy tight-binding model, derived from the LDA calculation, with atomic SOC in it,  they succeeded in obtaining an insulating solution for NaIrO$_3$ for reasonable values of SOC ($\lambda \approx 0.33$ eV) and $U=$2 eV, relevant for NaIrO$_3$. Furthermore, they identified the insulating state of NaIrO$_3$ as a ``band insulator" rather than a Mott insulator [see Fig. \ref{fig9} (d)]. 
 
Ref. \cite{Bhowal2015} further emphasized the influence of the structural geometry in driving the insulating state in NaIrO$_3$. Structural optimization, keeping the experimental symmetry and the 
lattice parameters intact, showed that the position of the light oxygen atoms were changed significantly (see Table \ref{tab2}), leading to an almost insulating state, as depicted in Fig. \ref{fig9} (e). This further supports earlier finding of band insulating state in NaIrO$_3$, indicating correlation effects may not be significant in NaIrO$_3$.  With these new theoretical insights, it might be worthwhile to revisit the transport properties of NaIrO$_3$, possibly on single crystal samples, motivating future experiments in this direction.

\begin{table}
\caption{\label{tab2}
Comparison of the Wyckoff positions for the experimental and the optimized crystal structure of the primitive unit cell of NaIrO$_3$, as reported in Ref. \cite{Bhowal2015}.
} 
\begin{indented}
\lineup
\item[]\begin{tabular}{@{}*{7}{l}}
\br 
Atoms \ & \multicolumn{3}{c} {Experimental structure} \ & \multicolumn{3}{c} {Optimized structure} \cr
 \mr                           

\ & x \ & y \ & z \ & x \ & y \ & z\cr
 \br
 Na \ & 0.0 \ & 0.2507 \ & 0.25 \ & 0.0 \ & 0.2513  \ & 0.25 \cr
Ir \ & 0.0 \ & 0.0  & 0.0 \ & 0.0 \ & 0.0 \ & 0.0  \\
O$_1$ \  & 0.5  \ & 0.43766 \ &  0.25 \ & 0.5  \ & 0.42621 \ & 0.25 \cr
O$_2$  \ & 0.5 \ & 0.10159 \ & 0.05681 \ & 0.5  \ & 0.11652 \ & 0.05852 \cr
\br          	 
\end{tabular}
\end{indented}
\end{table}

\subsection{6H perovskite iridates}

We now discuss the $d^4$ 6H-perovskite iridates with the general formula Ba$_3M$Ir$_2$O$_9$. The divalent cation $M$, e.g., Ca, Mg, Zn, Sr renders 5+ charge state to Ir.  These 
6-H perovskite iridates were synthesized by Sakamoto {\it et. al}~\cite{Sakamoto} back in 2006 and, then, further revisited with an idea to understand the  SOC effects  \cite{Nag2016,NagBMIO2018,Nag2019,Salman}.  It is interesting to point out here that similar to the double perovskite structure, 6H pervoskite structure also offers the composition flexibility, viz., the varying charge state of the $M$ cation results in different valency of the Ir ion, which includes the rather unusual fractional charge states in addition to the common
4+ charge state of Ir atom \cite{Sakamoto, Thumm,Doi2004,Dey2013,Panda2015,JChakraborty}. For example, Li ion at the $M$ site or trivalent $M$ ions such as, Y, Sc, In, give rise to fractional charge states 5.5+ ($d^{3.5}$), and 4.5+($d^{4.5}$) at the Ir site respectively, while tetravalent $M$ ions, e.g., Ti and Zr result in a 4+ charge state of Ir ion ($d^{5}$). 
Here we address the two key issues, viz., the impact of the local distortion, assisted by the $M$ cation, on the magnetic properties of the 6H perovskite iridates and, also, discuss the hopping induced formation of the magnetic moment at the Ir site.

\subsubsection{Structural distortion and its impact on the magnetic properties of $ Ba_3MIr_2O_9$:}

\begin{figure}[t]
\centering
 \includegraphics[width=\linewidth]{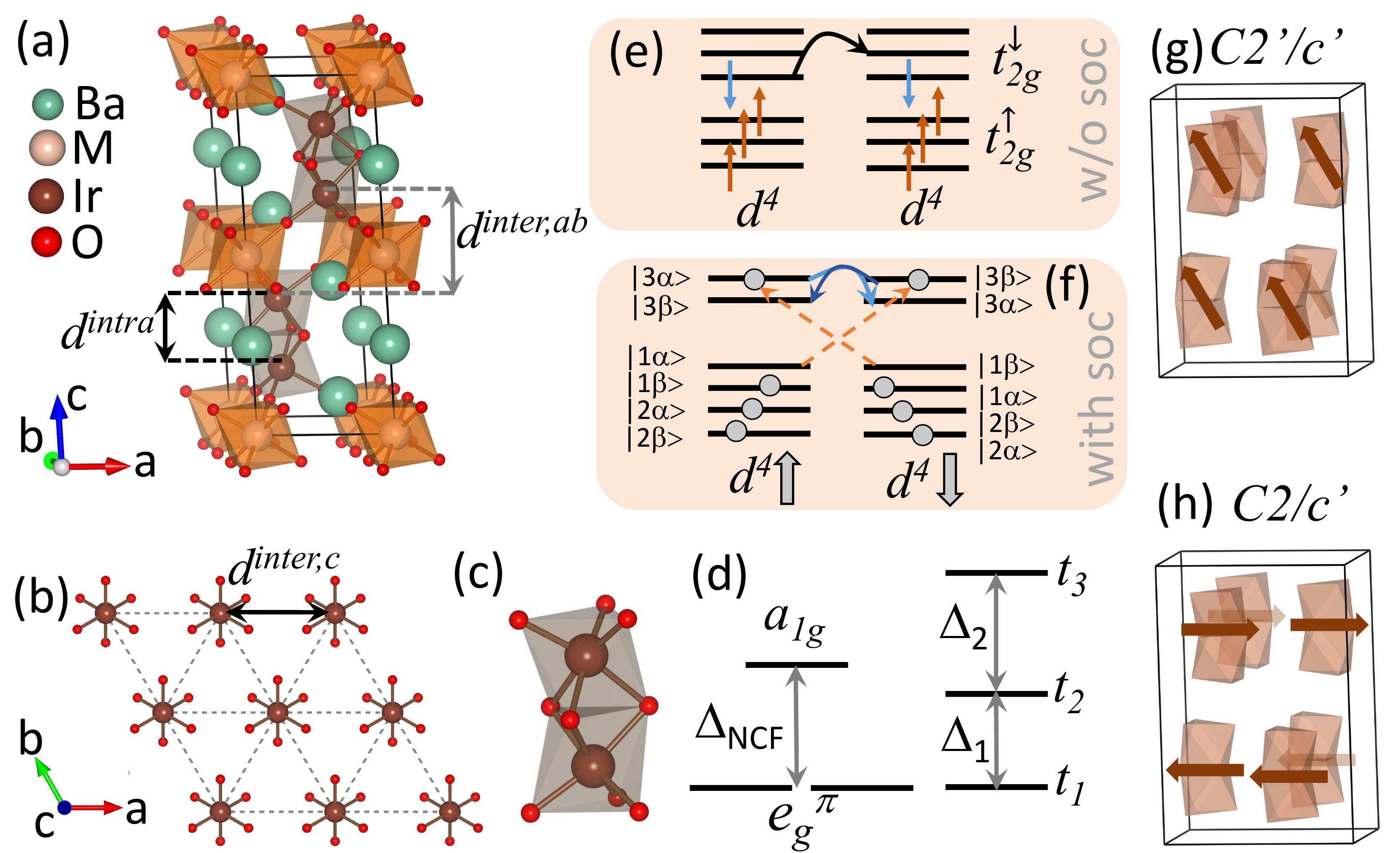}
 \caption{Structural distortion and magnetic properties of Ba$_3$MIr$_2$O$_9$. (a) Crystal structure of Ba$_3$MIr$_2$O$_9$. The structure consists of face-shared Ir$_2$O$_9$ units, which are arranged on a triangular network, shown in (b). The intra- ($d^{intra}$) and inter-dimer ($d^{inter,ab}$ and $d^{inter,c}$) distances are indicated in (a) and (b). (c) The Ir$_2$O$_9$ unit, that undergoes different types of distortion for $M=$ Zn, Mg, Ca, and Sr (see Table \ref{tab3}). This results in a trigonal crystal field splitting $\Delta_{\rm NCF}$ in Zn and Mg compounds ({\it left}), while the $t_{2g}$ degeneracy is completely removed for Sr and Ca compounds ({\it right}), shown schematically in (d). The non-degenerate $t_{2g}$ states for Sr and Ca are denoted by $t_1, t_2$, and $t_3$ and the energy separations between the consecutive levels are denoted by $\Delta_1$ and $\Delta_2$ respectively. The cartoons, adopted from ref. \cite{Nag2016}, showing (e) the FM and (f) the AFM interactions between Ir$^{5+}$ ions in absence and presence of SOC (see text for details).  
 (g) and (h) show the symmetry allowed magnetic structures corresponding to the magnetic space groups $C2'/c'$ and $C2/c'$ which are found to be energetically lowest magnetic configurations for Sr and Ca compounds respectively. Figures (g) and (h) are adopted from Ref. \cite{NagBMIO2018}.}
\label{fig10}
\end{figure}

The basic crystal structure of Ba$_3$MIr$_2$O$_9$ is quite intriguing. In contrast to the previously discussed structures,  here the IrO$_6$ octahedra share their faces with each other forming thereby  Ir$_2$O$_9$ bi-octahedra, which in turn leads to the formation of structural Ir-Ir dimer. Such dimers are again arranged on a geometrically frustrated triangular network, as illustrated in Fig. \ref{fig10}. 
Furthermore, the size of the cation $M$, has a significant impact on the crystal geometry, which, in turn, affect the magnetic properties in different members within the same 6H perovskite family. For example, the Rietveld analyses of the X-ray diffraction data \cite{Sakamoto,Nag2016,NagBMIO2018}  show that while Ba$_3$ZnIr$_2$O$_9$ and Ba$_3$MgIr$_2$O$_9$ with the smaller cations Zn and Mg crystallize in the hexagonal $P6_3/mmc$ space group, having less distorted IrO$_6$ octahedra, rest of the compounds, where the cation M is larger in size,  adopt the crystal structure with monoclinically distorted $C2/c$ space group, having significantly distorted IrO$_6$ octahedra.  While in the Zn and Mg compounds, the trigonal distortion of the face-shared IrO$_6$ octahedra in the hexagonal symmetry splits the Ir-$t_{2g}$ levels 
into a low-lying doublet ($e_g^{\pi}$) and a singlet ($a_{1g}$) with a non-cubic crystal field splitting $\Delta_{\rm NCF}$, the  
monoclinic symmetry of the Ca and Sr compounds leads  to complete removal of the
degeneracy of the $t_{2g}$ orbitals, denoted by $t_1, t_2,$ and $t_3$ in Fig. \ref{fig10} (d), with the corresponding energy differences $\Delta_1$ and $\Delta_2$ respectively (see Fig. \ref{fig10} (d) for pictorial representation). The stronger distortion in the latter compounds also leads to stronger non-cubic crystal field splitting in these systems. The comparison of the crystal geometry and the resulting non-cubic crystal field splitting ($\Delta_{\rm NCF}$ or $\Delta_1$, $\Delta_2$) are listed in Table \ref{tab3}.

The impact of the crystal geometry on the magnetic properties have been studied both theoretically and experimentally. While Sakamoto {\it et al} showed that all the systems behave paramagnetically with no long range magnetic order down to 1.8 K and described the magnetic properties of these systems by  Kotani’s theory~\cite{Kotani}, later a combined experimental and theoretical analysis~\cite{Nag2016} of Ba$_3$ZnIr$_2$O$_9$ showed that the compound exhibits a  fascinating  spin-orbital liquid state down to at least 100 mK. 
In contrast to the favorable FM interaction in absence of SOC between two Ir ions with $d^4$ electronic configuration, the spin-orbit entangled $\Gamma_7$ and $\Gamma_8$ states at the neighboring Ir sites, originating from the Ir-$t_{2g}$ orbitals in presence of SOC,  interact with each other anti-ferromagnetically [see Fig. \ref{fig10}(e) and (f)]. The key to the AFM interaction in presence of SOC is the unusual hopping between pesudo-spins. Unlike the hopping between real spin that are always spin-conserved, the spin-orbital entanglement also allows for certain hopping between two states with different pseudo-spins, e.g., $\Gamma_8$ electrons can virtually hop from $|1\alpha\rangle$ ($|1\beta\rangle$) to $|3\beta\rangle$ ($|3\alpha\rangle$) state. These electrons, which are promoted to the $\Gamma_7$ state further can gain energy by hopping to the other $\Gamma_7$ state as illustrated in Fig~\ref{fig10}(f). This later hopping is, however, forbidden if the interaction is ferromagnetic, thereby stabilizing an AFM interaction between the closely spaced Ir atoms forming dimers. Such AFM interaction manifests in the negative Curie-Weiss temperature $\theta_{\rm CW}\sim$ -30 K with an effective moment $\sim$0.2 $\mu_B$/Ir in the susceptibility measurements.  The magnetic moment at the Ir site represents the hopping assisted excitation from $J=0$ state to $J=1$ state. We come to the detail discussion of this point in the latter part of this section. 

\begin{table}
\caption{\label{tab3}
The comparison of the structural parameters and the resulting non-cubic crystal field splitting in Ba$_3M$Ir$_2$O$_9$, $M =$Zn, Mg, Ca and Sr.
} 
\begin{indented}
\lineup
\item[]\begin{tabular}{@{}*{5}{l}}
\br 
 \ & Ba$_3$ZnIr$_2$O$_9$ \ & Ba$_3$MgIr$_2$O$_9$  \ & Ba$_3$CaIr$_2$O$_9$ \ &Ba$_3$SrIr$_2$O$_9$          
 \cr
 \mr              
a (\AA)\ &   5.77034 \ & 5.7705  \ &  5.9091   \ &  5.9979 \cr
b (\AA) \ &  5.77034  \ &   5.7705             \ & 10.2370 \ &  10.3278 \cr
c (\AA)\ &  14.3441 \ & 14.3216 \ & 14.7624 \ & 15.1621 \cr
$\beta$ ($^{\circ}$)  \ &    \ &              \ & 91.14 \ & 92.78 \cr  
Ir-O1 (\AA)              \ &   2.021 \ & 2.058  \ & 2.010 \ & 2.041 \cr                                                                                                                   
Ir-O2 (\AA)                \ &   1.948  \  &        1.932 \  & 2.079 \ & 2.026, 2.181 \cr                                                                                                                
Ir-O3 (\AA)                  \ &            \ &                      \ & 1.849 \ & 1.991 \cr
Ir-O4 (\AA)                      \ &            \   &                 \ & 1.865 \ & 1.839 \cr
Ir-O5 (\AA)                       \ &           \ &                  \ & 2.092 \ & 2.079 \cr                                                             
$d^{intra}_{\rm Ir}$ (\AA)         \ &     2.75    \ & 2.76 \ & 2.735\ & 2.69 \cr                                                                                                                   
$d^{inter,c}_{\rm Ir}$ (\AA) \ &     5.54    \ &   5.52  \ &  5.52, 5.78, 5.86 \ & 5.86, 5.99, 6.11 \cr                                                                                                                    
$d^{inter,ab}_{\rm Ir}$ (\AA)I      \ & 5.78 \ & 5.77  \ & 5.909, 5.91,  5.91 \ & 5.97, 5.97, 5.99 \cr
$\angle$Ir-O1-Ir ($^{\circ}$)                              \ & 85.92 \ & 85.96 \ & 85.75 \ & 82.55\cr                                                                                                                    
$\angle$Ir-O2-Ir  ($^{\circ}$)                             \ & \ & \ & 80.06 \ & 79.50 \cr
$\Delta_{\rm NCF}$ (eV)            \ &    0.04      \ &   0.06   \ & 0.07, 0.24  \ & 0.09, 0.16  \cr
\br
\end{tabular}
\end{indented}
\end{table}

The AFM interaction within the Ir-Ir dimer leads to the formation of spin-orbital singlet (SOS) state. Each of these SOSs, further, anti-ferromagnetically interact with six other neighboring SOSs located on a frustrated triangular network within the $a-b$ plane and three out of the plane neighbors, corresponding to the Ir-Ir distances $d^{inter,ab}_{\rm Ir}$ and $d^{inter,c}_{\rm Ir}$ in Table \ref{tab3} respectively. The energy scales of these intra-dimer $J^{intra}_{d}$ (-14.6 meV, with neighboring single Ir ion) and inter-dimer interactions  $J^{inter}_{d}$ (-1.5 meV for six neighbors within the $ab$ plane and -1.6 meV with three neighbors perpendicular to the $ab$ plane) being comparable, a  strong magnetic frustration is developed in the system, giving rise to a spin-orbital liquid state, as evident from the muon-spin rotation, magnetization and heat capacity measurements \cite{Nag2016}.

The magnetic properties of the Mg compound is very similar to the Zn compound, showing noticeable AFM interaction from magnetic susceptibility measurement, although the heat capacity and $\mu$SR measurements suggest a weaker frustration in this case. The weaker frustration in the Mg compound results from the stronger $J^{intra}_{d}$ (-21.3 meV) compared to $J^{inter}_{d}$ (-0.9 meV within the $a-b$ plane and -1.4 meV out-of-plane interaction) in the system. On the other hand, the strong distortion of the IrO$_6$ octahedra in the Sr and the Ca compounds break the local inversion symmetry, giving rise to uncompensated nearest neighbor Dzyaloshinskii-Moriya (DM) interactions. These DM interactions are expected to cause the canting of the spins, resulting in non-collinear magnetic structures. 
Indeed, explicit DFT calculations \cite{NagBMIO2018} with the symmetry allowed magnetic structures corresponding to an assumed propagation vector $\vec k = (0,0,0)$ shows that the lowest energy magnetic configuration corresponds to the magnetic space group $C2/c'$, in which the Ir ions forming the dimer have anti-parallel spin components along the $x$, and $z$ directions while due to canting have  a parallel $y$ component.  This net $y$ component of spin moments from a pair of neighboring Ir ions, however, is anti-parallel to the the net $y$ moment of another pair of Ir ions, situated along the $c$ direction, leading to a net zero moment in the system. 
The weak dimeric feature observed in the magnetic susceptibility of Ba$_3$CaIr$_2$O$_9$ may be an outcome of the antiparallel inter-dimer arrangement of moments in the $C2/c'$ magnetic structure as shown in Fig. \ref{fig10}(g). On the contrary, similar analysis \cite{NagBMIO2018} showed that another symmetry allowed magnetic configuration $C2'/c'$ is energetically favorable  for the Sr compound. Interestingly, this magnetic configuration allows for a net FM moment in the system, possibly leading to a ferro-magnetic like feature in magnetic
susceptibility under low magnetic fields. Thus, the family Ba$_3$MIr$_2$O$_9$ with $M =$ Zn, Mg, Ca, Sr constitutes an intriguing platform to investigate the effect of local structural distortion  in the magnetic properties of spin-orbit entangled $d^4$ state.

\subsubsection{Hopping induced magnetism:}

\begin{figure}[t]
\centering
 \includegraphics[width=\linewidth]{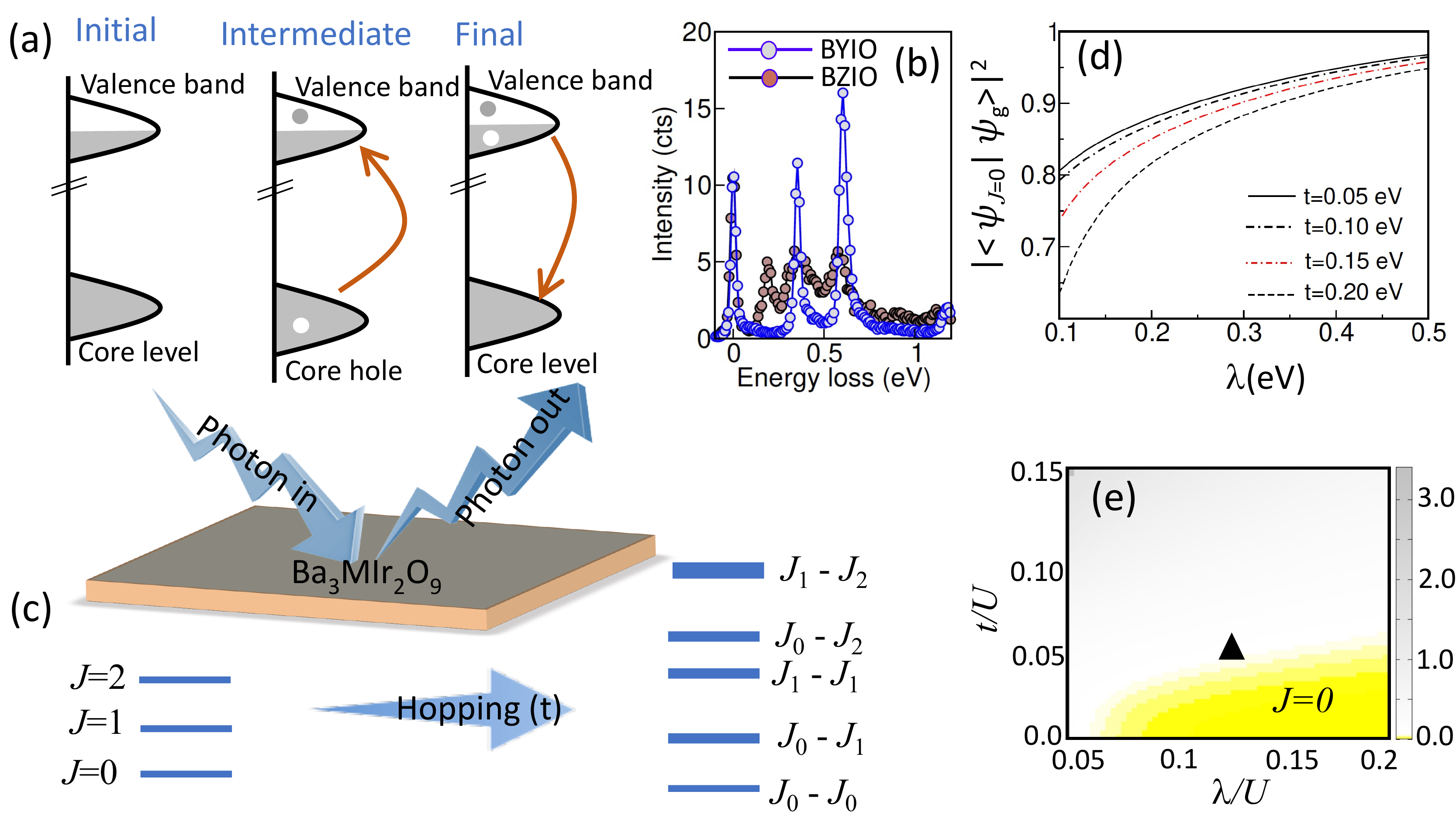}
 \caption{The role of hopping in the RIXS spectrum of the 6H iridates. (a) Schematic illustration of the RIXS process in general. (b) The low-energy RIXS spectra from the SOC states in Ba$_3$ZnIr$_2$O$_9$ (BZIO). In marked contrast to the corresponding spectra in Ba$_2$YIrO$_6$ (BYIO), the spectra in 6H perovskite depicts the presence of additional inelastic peaks and spectral-weight transfer to lower energies. 
  (c) The energy states obtained from an atomic model and a two-site model. The higher energy states with low intensity are not shown. The feature of the observed spectra for 6H iridates in (b),  although can not be explained within an atomic model but is a characteristic of a two-site model indicating the crucial role of hopping in the present iridates. (d) The deviation of the ground state $\psi_g$ of a two-site model, relevant for 6H iridates,  from the $J=0$ state. 
   (e) The same can be concluded from the phase diagram, showing the variation of the expectation value of $J^2$ operator in the parameter space of $t/U$ and $\lambda/U$. The black triangle in the phase diagram corresponds to the parameters $t$ = 0.1 eV, $\Delta_{\rm NCF}$ = 0.02 eV, $J_{\rm H}$ = 0.5 eV, $\lambda$ = 0.25 eV and $U$ = 2.0 eV, relevant for the iridates.
}
\label{fig11}
\end{figure}

As already discussed, a spontaneous magnetic moment, due to the large Ir-Ir hopping, is developed at the Ir site in Ba$_3$MIr$_2$O$_9$ ($M =$ Zn, Mg, Ca, Sr). In contrast to the double perovskite iridates, where the Ir atoms are separated apart, in the 6H perovskite structure the Ir atoms form dimer, which constitute the basic building block of the 6H structure. This strong Ir-Ir dimer interaction manifests itself in the RIXS spectrum, that probes the low energy excitations of these iridates. In contrast to the double perovskite iridates, the RIXS spectrum in Ba$_3$MIr$_2$O$_9$  has distinct features, as shown in Fig. \ref{fig11} (b), which can not be captured within an atomic model. A meaningful description of such RIXS spectrum requires at least a two-site model with finite inter-site hopping, emphasizing the importance of the band structure effect on the ground state properties of these series of 6H iridates. Here, we provide an overview of the band structure effect and its implications to the RIXS spectra observed for  Ba$_3$MIr$_2$O$_9$ systems.

We begin with a brief description of the RIXS process. 
RIXS is a second order ``photon-in photon-out'' spectroscopic process, as illustrated schematically in Fig. \ref{fig11} (a). In case of $d^4$ iridates, the incoming photon with frequency $\omega_1$ excites the Ir-$2p$ core electron to the Ir-$d$, $t_{2g}$ unoccupied states (corresponding to the L$_3$ absorption edge) forming an intermediate state $\vert n\rangle = \vert 2p_{\frac{3}{2}}^{\rm hole} t_{2g}^5\rangle$ with core-hole. To fill up the hole in the core level, a different $t_{2g}$ electron corresponding to the initial $t_{2g}^4$ configuration may come back to the $2p$ state resulting in an emission of a photon of frequency $\omega_2$ and a final state $\vert f\rangle=\vert t_{2g}^4\rangle$.  Note that the final state $\vert f\rangle$ can be any kind of arrangement of four electrons within the three $t_{2g}$ orbitals and the energy difference $\omega_1-\omega_2$ corresponds to these {\it d}-{\it d} excitation energies. The estimation of these energies are, therefore, beyond the scope of single-particle description and it is important to consider a many-body Hamiltonian with proper interactions.

The typical RIXS spectrum for Ba$_3$MIr$_2$O$_9$ is shown in Fig. \ref{fig11} (b). Such a spectrum, in particular the additional peaks as shown in the figure, can not be described by the triplet ($J=0 \rightarrow J=1$) and quintet ($J=0 \rightarrow J=2$) spin-orbit excitons of an atomic model for spin-orbit coupled $d^4$ systems. Indeed, the two-peak feature is  the characteristic of a two-site model with a finite inter-site hopping [see Fig. \ref{fig11} (c)], indicating the necessity to go beyond the atomic limit to incorporate the non-local effects, such as non-cubic crystal field and the hopping, present within the solid.

The spin-orbit exciton energies in the RIXS spectra may be mapped into the corresponding differences in the energy eigenvalues computed using exact diagonalization of the  two-site model (similar to the Hamiltonian discussed in section \ref{model2}, Eq. \ref{two-site} with an additional non-cubic crystal field term). 
In the two-site model, the atomic states interact with each other in presence of significant hopping giving rise to a new set of spin-orbit coupled states, e.g., atomic states $J=1$ ($J_1$) at each sites interact with each other to give rise to a new set of nine states, which we denote here by $J_1-J_1$ for simplicity.  The resulting new sets of spin-orbit coupled states are: 
$J_0-J_0$ (1), $J_0-J_1$ (6), $J_1-J_1$ (9), $J_0-J_2$ (10), $J_1-J_2$ (30), $J_2-J_2$ (25), etc, where the number within the parenthesis indicates the number of states arising due to interaction between the two-atomic states. 
 The excitation from the $J_0-J_0$ state to the states $J_1-J_1$ and $J_0-J_2$ states  describe the broad peak around $\sim 0.4$ eV in the RIXS spectrum, observed in the experiments. 

An estimation of the various parameters, starting from a realistic set of values extracted from the {\it ab initio} calculations, may be obtained when the theoretically computed excitation energies coincide with the extracted energy loss values within the uncertainty of the measurement (full width at half maximum of the Lorentzian function). Note that unlike the atomic model, the energies of the spin-orbit excitons obtained from the two-site model depend on the Coulomb repulsion $U$ possibly due to the superexchange energy scale $\sim t^2/U$ involved in the latter case.  The estimated Hund's coupling $J_H$ is about $\sim 0.45$ eV, which is comparable to the other $5d^5$ materials but higher than the $d^4$ double perovskite iridates \cite{Yuan,Kusch}.  More interestingly, the SOC $\lambda$ is highly suppressed to a value of $\sim 0.26$ eV, highlighting the decisive role of hopping in these  6H iridates. The large intra-dimer hopping and the strongly suppressed SOC leads to deviation of the two-site ground state from the ideal $J=0$ state by about 10$\%$, resulting in a non-zero moment as also evident from the ground state expectation value of $\langle J^2 \rangle$ [see Fig. \ref{fig11} (d) and (e)].  In a nut shell, Ba$_3$MIr$_2$O$_9$ is an exemplary $d^4$ system, where the hopping dramatically affects the spin-orbit coupled  ground state, leading to formation of moment at the ``atomically non-magnetic" Ir site.

\subsection{Hexagonal Iridates Sr$_3$MIrO$_6$}

\begin{figure}[t]
\centering
 \includegraphics[scale=.3]{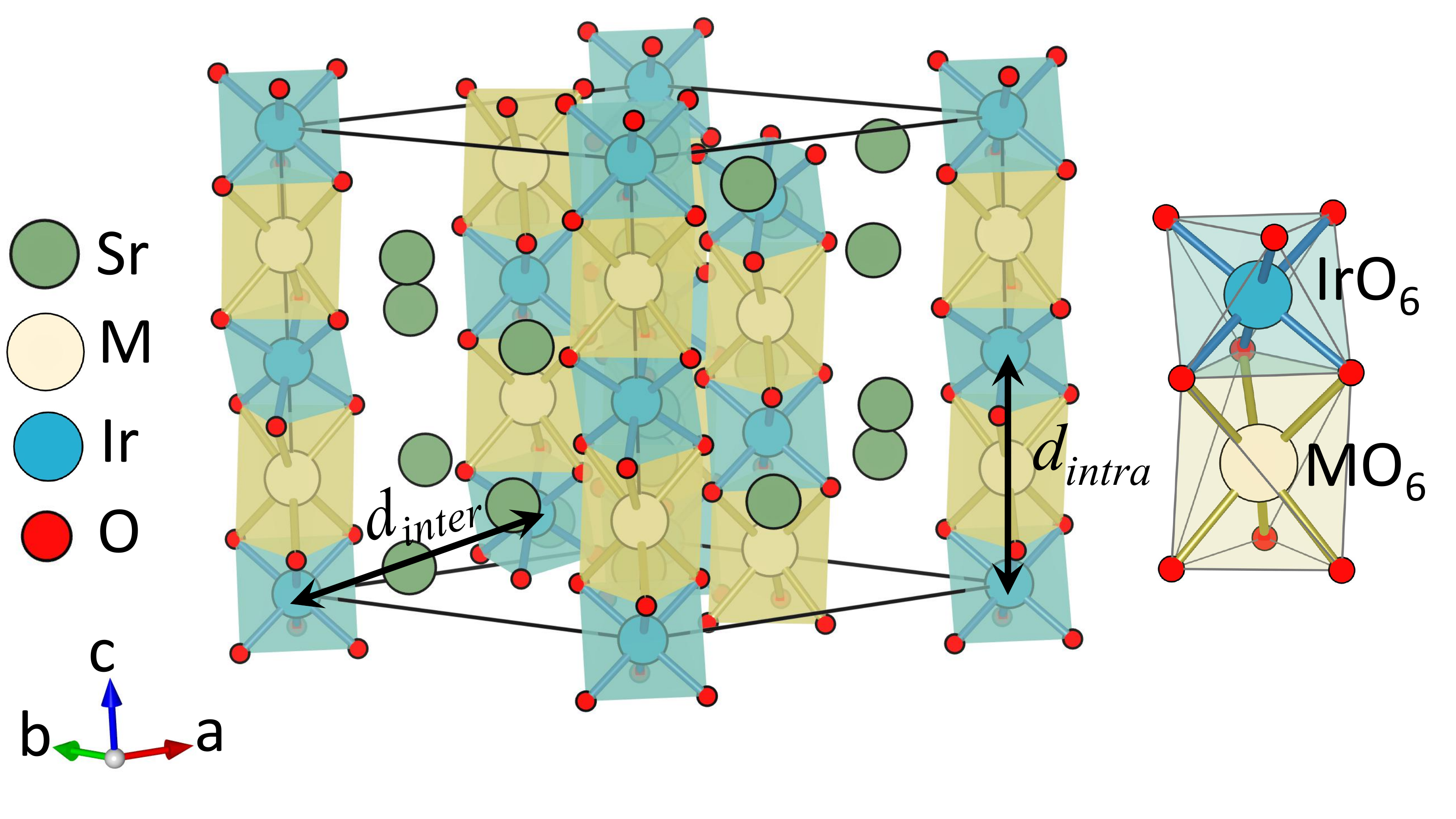}
 \caption{ Crystal structure of Sr$_3$MIrO$_6$, showing the alternating chain of IrO$_6$-SrO$_6$ running along the crystallographic $c$-axis. The face-shared IrO$_6$ and MO$_6$ unit is shown separately. The Ir atoms, both within the chain ($d_{intra}$), and between the chains ($d_{inter}$), are quite well separated in the structure, which leads to narrow Ir-$t_{2g}$ bands.
}
\label{fig12}
\end{figure}

The pentavalent iridate Sr$_3$MIrO$_6$ crystallizes in the hexagonal structure with the $R\bar{3}c$ space group symmetry, which is isostructural to the parent compound Sr$_4$IrO$_6$. The doping with alkaline metals $M$, e.g., Na, Li, K at the Sr site results in a reduced charge state of Ir, viz., the pentavalent iridate, the subject of the present review. 
To the best of our knowledge these pentavalent iridates have been synthesized as early as 1996 \cite{Segal,Davis,Frenzen1996}. The crystal structure of Sr$_3$MIrO$_6$ is quite interesting in the sense that the Ir atoms are far apart from each other, constituting isolated IrO$_6$ octahedral unit. These octahedra share their face with the neighboring MO$_6$ trigonal prism, forming thereby an alternating chain of IrO$_6$-SrO$_6$ along the crystallographic $c$-axis. The crystal structure of  Sr$_3$MIrO$_6$ consists of such well separated alternating chains, as shown in Fig. \ref{fig12}. The  large Ir-Ir distance and the lower connectivity of Ir atoms results in a narrow Ir-$t_{2g}$ bands, which, in turn, gives rise to the hope of realizing the $J=0$ state. Interestingly, existing experiments \cite{Segal,Davis} indicate a temperature-independent paramagnetic behavior for Sr$_3$MIrO$_6$, with $M =$Na, Li, which further points towards a non-magnetic $J=0$ state.

With this viewpoint, recently DFT calculations \cite{Ming} have been performed on both Na and Li compounds. These calculations find a non-magnetic state for these two systems. With the projection on to the $j_{\rm eff}$ states, the authors show that the $j_{\rm eff} =3/2$ and  $j_{\rm eff} =1/2$ states are very well separated in energy without any mixing between them. Interestingly, the $J=0$ states, proposed for this system, are insulating in nature even without Coulomb correlation. This is in sharp contrast to the other $d^4$ iridates, discussed previously in this review, where finite Coulomb correlation is always required to open up an energy gap. The insulating property of these systems are, therefore, better described as SOC driven band insulating state, rather than the Mott insulating state.
A simple explanation to this behavior lies within the reduced dimensionality of the structure, resulted from the isolated IrO$_6$ octahedra, as discussed before. As result of this structural low dimensionality, the bandwidth also reduces which, in turn, favors an insulating state.
 Notably, the bandwidth of these materials are almost half of the Ir-$t_{2g}$ bandwidths found in the double perovskite and 6H perovskite structures which may be conducive for the realization of the $J=0$ state. 

Interestingly, recent experiments \cite{Ray} however point towards magnetism at the Ir site for both Na and Li compounds, with even larger effective magnetic moment. Notably, the IrO$_6$ octahedra in these systems are not ideal and undergo distortions. 
It might be, therefore, interesting to revisit these materials both experimentally and theoretically with the new insights at hand to confirm or refute the presence of $J=0$ non-magnetic state in Sr$_3$MIrO$_6$.  
It is important to point out here that the isostructural $d^5$ sister compound Sr$_3$NiIrO$_6$ exhibits a strong single-ion magnetic anisotropy \cite{Birol}.  In presence of any magnetism in the $d^4$ counterparts, it would be furthermore interesting to see whether  such single-ion magnetic anisotropy is also present in pentavalent iridates.

\section{Outlook}

Understanding electronic and magnetic properties of iridates is a rapidly growing field of research, a part of which are pentavalent iridates. While most of these pentavalent iridates have been synthesized long ago, e.g., the first $d^4$ Ir-based material was synthesized as early as 1960's \cite{Figgis,Earnshaw}, understanding and revealing many of the interesting properties of these materials remain elusive until recently. The surge in the interest of discovering  intriguing properties of $d^4$ iridates are primarily driven by the theoretical proposal of excitonic magnetism in these materials. While theoretical and experimental understanding of the ground state of real pentavalent iridate systems are not always conclusive, the enormous curiosity to understand the electronic and magnetic properties of these materials over the past decade builds up the solid ground to further tune and manipulate the states to get desired properties of matter. While the review of each of the members of this giant family of materials is beyond the scope of the present article, the present review provides an overall view of this family, highlighting the important issues that have been of interest to the scientific community and focussing on a few specific materials that have received considerable attention over the time.

It is hard to draw any conclusion to the iridates research as the field is still at its peak with more than 300 publications over the past five years (according to the ``iridates" keyword search in the web of science) and many more avenues to be explored in future.  
We, therefore, end this review by drawing the attention of the readers to a few interesting open questions or  directions that still need to be  answered or ventured.    

The ground state properties of $d^4$ double perovskite iridates have been at the centre of debate and controversies. The Ir atoms do not order magnetically in most of the $d^4$ iridates, including the double perovskite systems. A more clear and better understanding of the ground states of these systems may be obtained by studying the magnetic properties of these systems under pressure or strain. Intuitively speaking, the tensile strain would separate the Ir ions far apart, leading to more atomic like scenario, where the spin-orbit effects prevail. On the other hand, the compressive strain would place the Ir ions close to each other, and if the strain is strong enough a competition between SOC and covalency may result in. This, in turn, may lead to the onset of long range order. It might be interesting to see the effect of strain on the spin-orbit excitons as well, which may be probed in RIXS experiments. The spin-orbit excitons have received a great deal of attention in recent times, particularly driven by the fact that its condensates near phase transitions exhibit dispersive Higgs mode, a long-sought-after physical
phenomenon  \cite{Jain2017}. External perturbations such as pressure or strain may trigger this critical transition points, that results from the competition between various non-local effects and spin-orbit effect, which may be difficult to attain intrinsically in the bulk iridates.

The strain effects might also shed light into the insulating nature of the iridates. While most of the iridates are found to be insulating, their origin of the insulating behavior, as discussed in the literature, is actually quite different. For example, while NaIrO$_3$ and Sr$_3$MIrO$_6$ are close to  band insulating state \cite{Du,Ming}, double perovskite iridate Ba$_2$YIrO$_6$ is described as a Mott insulator \cite{TDey}. It is noteworthy that Ref. \cite{Paramekanti} also indicated possible connection between the AFM configuration and the insulating state. Structural modifications with pressure or strain might be useful to understand and analyze the transport properties of the different $d^4$ iridates, revealing also the connection between magnetism and insulating state, if any. It is interesting to point out here that the tetravalent perovskite iridate exhibits dramatical changes in the electrical conducting properties in the two-dimensional limit \cite{Schutz,Groenendijk}. The corresponding behavior of the pentavalent counterparts may be worth investigating. 

While the magnetic interactions in the $d^4$ iridates are either very weak or completely absent, it would also be interesting to look for higher order ``hidden" parameter, magnetic and/or non-magnetic, which may act as primary order parameters, governing the properties of the materials. Since SOC, in general, is often found to promote such hidden orders,  $d^4$ iridates are excellent materials to investigate them. Interestingly, such hidden orders have been predicted and also experimentally verified for $5d^2$ double perovskites \cite{Chen2011,Lovesey2020}. Since the $d^4$ electronic configuration is particle-hole analogue of $d^2$ configuration (two electrons in $t_{2g}$ orbitals are equivalent to four holes in the $t_{2g}$ orbitals), it would be interesting to explore the studied  pentavalent iridates from this point of view. 

Undoubtedly there are still plenty of avenues to be explored both theoretically and experimentally, motivating many more future researches on the pentavalent iridates in the coming years.

\section*{Acknowledgements}

We take this opportunity to  deeply acknowledge
 Avinash V. Mahajan, Sugata Ray, Tanusri Saha-Dasgupta,  and Santu Baidya, Atasi Chakraborty,  Shreemoyee Ganguly, Abhishek Nag, and Swarup K. Panda, who provided us with new insights and inspiring ideas and fruitful collaborations. 
We thank
all our collaborators in the iridate research over the last few years:
 Fabrice Bert, Pabitra K. Biswas, I. Carlomagno, Anna Delin, Anna Efimenko, Olle Eriksson, Paul G. Freeman, Adrian D. Hillier, Mitsuru Itoh, Som Datta Kaushik, Ying Li, Martin Mansson, Roland Mathieu, Philippe Mendels, Carlo Meneghini, Srimanta Middey,  Larse Nordstr\"om, Jean-Christophe Orain, Henrik M. Ronnow,  Marco M. Sala, Tapati Sarkar, Dipankar Das Sarma, Denis Sheptyakov, Vasudeva Siruguri, Mark T. F. Telling, Roser Valent\'i.  
 We also thank Bernhard Keimer, Giniyat  Khaliullin, Daniel I. Khomskii, Arun Paramekanti,  Sashi Satpathy, and Nandini Trivedi for many useful discussions. I. D. thanks Science and Engineering Research Board (SERB), India (Project No.EMR/2016/005925) and Department of Science and Technology-Technical  Research  Centre  (DST-TRC)  for financial support.

\section*{Appendix A: $l_{\rm eff} = -1$ state}

In presence of strong octahedral crystal field, atomic Ir-$d$ levels are split into three-fold degenerate $t_{2g}$  and an $e_g$ doublet. 
Now by writing $L_x~=~\frac{1}{2}(L_++L_-)$ and $L_y~=~\frac{1}{2i}(L_+-L_-)$ in terms of ladder operators and following the orbital momentum
algebra,
\begin{eqnarray}
  L_z|l,m_l\rangle& = &m_l|l,m_l\rangle \\ \nonumber
  L_{\pm} |l,m_l\rangle & = &\sqrt{l(l+1)-m_l(m_l\pm1)}|l,m_l\pm1 \rangle
\end{eqnarray}
we can calculate the matrix elements of $\langle t_{2g}|{\bf L}|t_{2g} \rangle$,
\begin{eqnarray}     \label{Lt2g}  
L_x  =\left( \begin{array}{ccc}
        0 & 0 & 0 \\		
        0 & 0 & i \\		
        0 & -i & 0 
        \end{array} \right)
L_y  = \left( \begin{array}{ccc}
       0 & 0 & -i\\
       0 & 0 & 0\\
       i & 0 & 0       
       \end{array} \right)
L_z  = \left( \begin{array}{ccc}
                 0 & i & 0\\
                 -i & 0 & 0\\
                 0 & 0 & 0 
                \end{array}  \right),
\end{eqnarray} 
where $t_{2g} \equiv \{ d_{yz}, d_{xz}, d_{xy}\}$.
Note that here we have dropped out the $\langle t_{2g}|{\bf L}|e_{g} \rangle$ terms, assuming that the $t_{2g}$ states are completely decoupled from the $e_g$ states, due to large separation in the energy scale, determined by $\Delta$. In reality, however, the $t_{2g}-e_g$ mixing may not be ignored. As discussed in Ref. \cite{Stamokostas}, the mixing may be crucial in interpreting the branching ratio, measured in experiments as well as in theory to construct effective models for the purpose of studying the magnetic phases and magnetic excitations. 
Nevertheless, if such mixing be ignored,  the projection of the {\bf L} operator 
for the $d$-electrons on the $t_{2g}$ manifold, as shown in Eq. \ref{Lt2g} may be identified as {\bf L}$(t_{2g})$~=~-{\bf L}$(p)$, where {\bf L}$(p)$ is the matrix representation of the {\bf L} operator in the basis of $p \equiv \{ p_x,
p_y , p_z \}$, indicating that the  
 $t_{2g}$ orbitals effectively behave like $p$-orbitals having angular momentum $l_{\rm eff}$~=-1, but with an extra negative sign. 



\begin{thebibliography}{10}
\bibitem{note} For hydrogenic systems, the SOC constant $\lambda$ strongly increases with the atomic number Z, $\lambda \propto Z^4$. 
However, alternative treatment (Landau and Lifshitz, 1965) shows that for the outer electrons, that  are responsible for the properties in the solid, 
$\lambda \propto Z^2$\cite{LL,Shanavas2014}. 
Both these dependences though are only
rough guides and the value of SOC in a real material can not be determined by any of these treatment, they however emphasize the strong dependence of  $\lambda$ on the atomic number Z.
\end{thebibliography}
\end{document}